\documentclass[12pt,preprint]{aastex}

\usepackage[dvipdfmx]{color}
\usepackage{amsmath}
\usepackage{booktabs}
\usepackage{natbib}
\usepackage{color}
\usepackage{ulem}

\newcommand{\myemail}{kitagawa.naomasa@gmail.com}
\newcommand{\kmpers}{\mathrm{km} \, \mathrm{s}^{-1}}
\newcommand{\logt}{\log T \, [\mathrm{K}]}

\slugcomment{}

\shorttitle{Doppler shift of the quiet region}
\shortauthors{Kitagawa, Hara, \& Yokoyama}

\begin{document}

\title{Doppler shift of the quiet region measured by meridional scans 
with the EUV Imaging Spectrometer onboard \textit{Hinode}}

\author{N.\ Kitagawa\altaffilmark{1}}
\affil{The University of Tokyo, 7-3-1 Hongo, Bunkyo-ku, Tokyo, 113-0033 Japan}

\author{H.\ Hara}
\affil{National Astronomical Observatory of Japan, 
2-21-1 Osawa, Mitaka-shi, Tokyo, 181-8588 Japan}

\and

\author{T.\ Yokoyama}
\affil{The University of Tokyo, 7-3-1 Hongo, Bunkyo-ku, Tokyo, 113-0033 Japan}

\altaffiltext{1}{e-mail: \myemail}

\begin{abstract}
Spatially averaged ($> 50''$) 
EUV spectral lines in the transition
region of solar quiet regions are known to be redshifted.
Because the mechanism underlying this phenomenon is unclear,
we require additional physical information on the lower
corona for limiting the theoretical models. To acquire this
information, we measured the Doppler shifts over a wide coronal 
temperature range 
($\logt=5.7$--$6.3$) using the spectroscopic data taken 
by the  \textit{Hinode} EUV Imaging Spectrometer.
By analyzing the data over the center-to-limb variations
covering the meridian
from the south to the north pole, we successfully measured the velocity 
to an accuracy of $3 \, \kmpers$.  
Below $\logt = 6.0$, the Doppler shifts of the emission lines were
almost zero with an error of $1$--$3 \, \kmpers$; above 
this temperature, they were blueshifted with a gradually increasing magnitude,
reaching $- 6.3 \pm 2.1 \, \kmpers$ at $\logt=6.25$.
\end{abstract}


\section{Introduction}
  \label{sect:cal_itdn}

The emission lines formed in the transition region
of the solar quiet regions
show redshifted features \citep[e.g.,][]{doschek1976,
brekke1997,chae1998doppler,peterjudge1999,teriaca1999}.
The magnitude of the redshift increases with increasing temperature, with
a peak of approximately  $10 \, \kmpers$ at $10^{5}\ \mathrm{K}$ 
and then decreases at higher temperatures. 
Using the Solar Ultraviolet Measurement of Emitted Radiation (SUMER) 
instruments 
of the \textit{Solar Heliospheric Observatory} (\textit{SoHO}) spacecraft,
\citet{peterjudge1999} found blueshifts for three coronal lines
%\textcolor{red}{
($-(2.4$--$3)\pm 1 \,\kmpers$ for Ne \textsc{viii} 
and $-4.5\pm 3 \,\kmpers$ for Mg \textsc{x});
%}
while \citet{chae1998doppler} showed that the same lines are redshifted
%\textcolor{red}{
($+5.3 \pm 1 \,\kmpers$ for Ne \textsc{viii} and
$+3.8 \,\kmpers$ for Mg \textsc{x}).
%}
%
These observations suggest prevalent 
%\textcolor{red}{\sout{vertical}}
mass motion in the heights of the observed Doppler-shifted emission 
lines.
To understand this motion, precise measurements within an accuracy
of a few $\kmpers$ are required.

The EUV Imaging Spectrometer (EIS; \citealt{culhane2007}) onboard 
\textit{Hinode} \citep{kosugi2007} measures 
Doppler shifts in the solar corona and the transition 
region with an unprecedentedly high accuracy. The CCD pixel size of the EIS
detector corresponds to $22\,\mathrm{m\AA}$ 
%\textcolor{red}{
($33 \,\kmpers$ for 200 \AA\ line)
%}
.
By fitting
a model function to an observed
profile in a band window with more than ten pixels, 
the precision can be improved to within a few $\kmpers$ in a statistical sense.
Since EIS has no onboard system for absolute 
wavelength calibration, the line centroid wavelength 
in a quiet region of the field of view (FOV) 
is frequently designated the zero-velocity point. 

Measurements of Doppler shifts contain several uncertainties.
First, uncertainty exists in the rest wavelengths of 
some emission lines, whether theoretically or experimentally derived.  
The NIST\footnote{\texttt{www.nist.gov/pml/data/asd.cfm}} database shows that 
the rest wavelengths are precise to the order of $10^{-2}${\AA}
%\textcolor{red}{
($15 \,\kmpers$ for 200 \AA\ line)
%}
in most cases, slightly larger than our current requirement.
Second is the uncertainty in the instrument's observing conditions: 
The grating component displaces with the thermal 
environmental change of the \textit{Hinode} spacecraft,
thus causing drift of the spectral signals on the CCDs \citep{brown2007}.
The SolarSoftware package developed by \citet{kamio2010} 
is widely used to correct the wavelength scale in EIS analysis,
but residuals of $4$--$5 \, \kmpers$ remain
in the standard deviation.
Third is the choice of the zero-velocity reference.
Many analyses assume zero Doppler shift for 
the averaged profile in a quiet region
of the emission line Fe \textsc{xii} $195.12${\AA}.
%
%\textcolor{red}{
%\sout{However, this emission line is too weak for measurements 
%in quiet regions.}
However, this choice is simply an assumption 
without any concrete physical reasoning.
%}

To avoid the above mentioned difficulties, 
we measured the
center-to-limb behavior of the Doppler shift.
This calibration is especially useful for determining
the zero-velocity wavelength of each spectral line
\citep[e.g.,][]{roussel-dupre1982}.
Applying this approach,
\citet{peter1999} and \citet{peterjudge1999} 
obtained the redshift
features in the transition region lines and the blueshift tendency
of the coronal lines from SUMER data.
Combining the observations of different instruments 
(such as SUMER and EIS) 
is also useful when one of the instruments (in this case, SUMER) has 
%\textcolor{red}{
%\sout{absolute wavelength calibration}
a substantially small uncertainty
with a reliable calibration method, e.g., in case of SUMER, that is
by using chromospheric lines and its counterpart
well-determined telluric lines \citep{samain1991}.
%}
\citet{2011A&A...534A..90D} conducted this analysis and
obtained the Doppler shift of spectral lines formed between 1 and 2 MK
in the quiet corona for the first time.

In this paper, 
we analyze the EIS data covering
the meridian from the south to the north pole, 
and measure the Doppler shifts
from the center-to-limb behavior of each emission line.
Our measurements are precise to within $3 \, \kmpers$. 
By using this technique, not only relative but also absolute
calibration becomes possible 
irrespective of the uncertainties in the line database and 
in the instrumental thermal drift.
Besides improving the precision, the EIS observations can provide 
new information of the Doppler shift over higher temperature ranges.
These independent measurements with independent assumptions
can be compared with those
of \citet{2011A&A...534A..90D} for better diagnostics of the
Doppler signals.

The rest of this paper is organized as follows.
Section 2 describes the data and observational setup. Section 3 is 
devoted to our data reduction procedures. Results and discussion are
given in Section 4 and 5, respectively. The Appendix details the selection 
process of the emission lines and presents examples of their profiles.

\section{Observations}

\begin{figure}[!bp]
  \centering
  \includegraphics[width=10.3cm,clip]{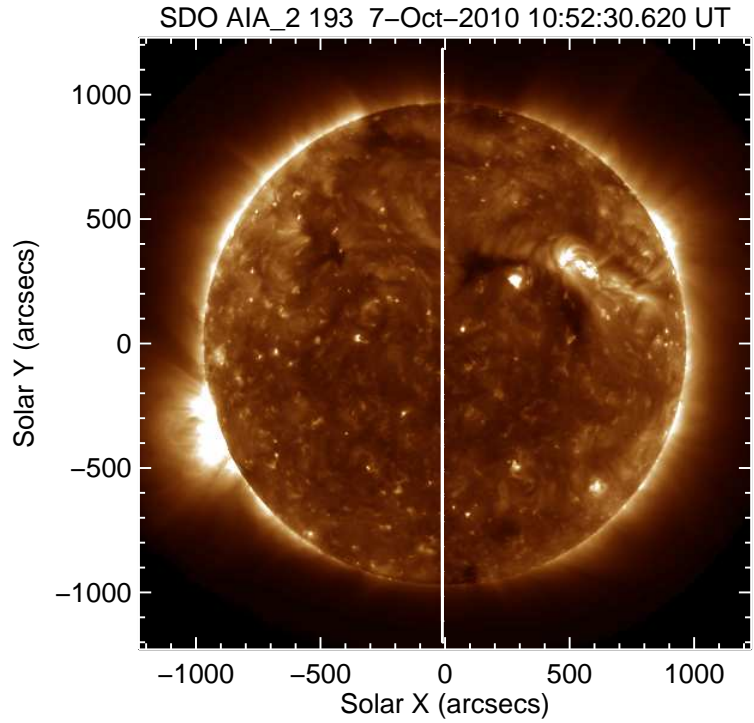}
  \caption{\textit{SDO}/AIA $193${\AA} passband image at the 
  start of the October sequence of HOP79. 
  The {white vertical} line indicates the location of spectral data 
  measurements by EIS.}
  \label{fig:data_hop79}
\end{figure}
\begin{figure}[!bp]
  \centering
  %bb=0 0 1471 536,
  \includegraphics[width=10.0cm,clip]{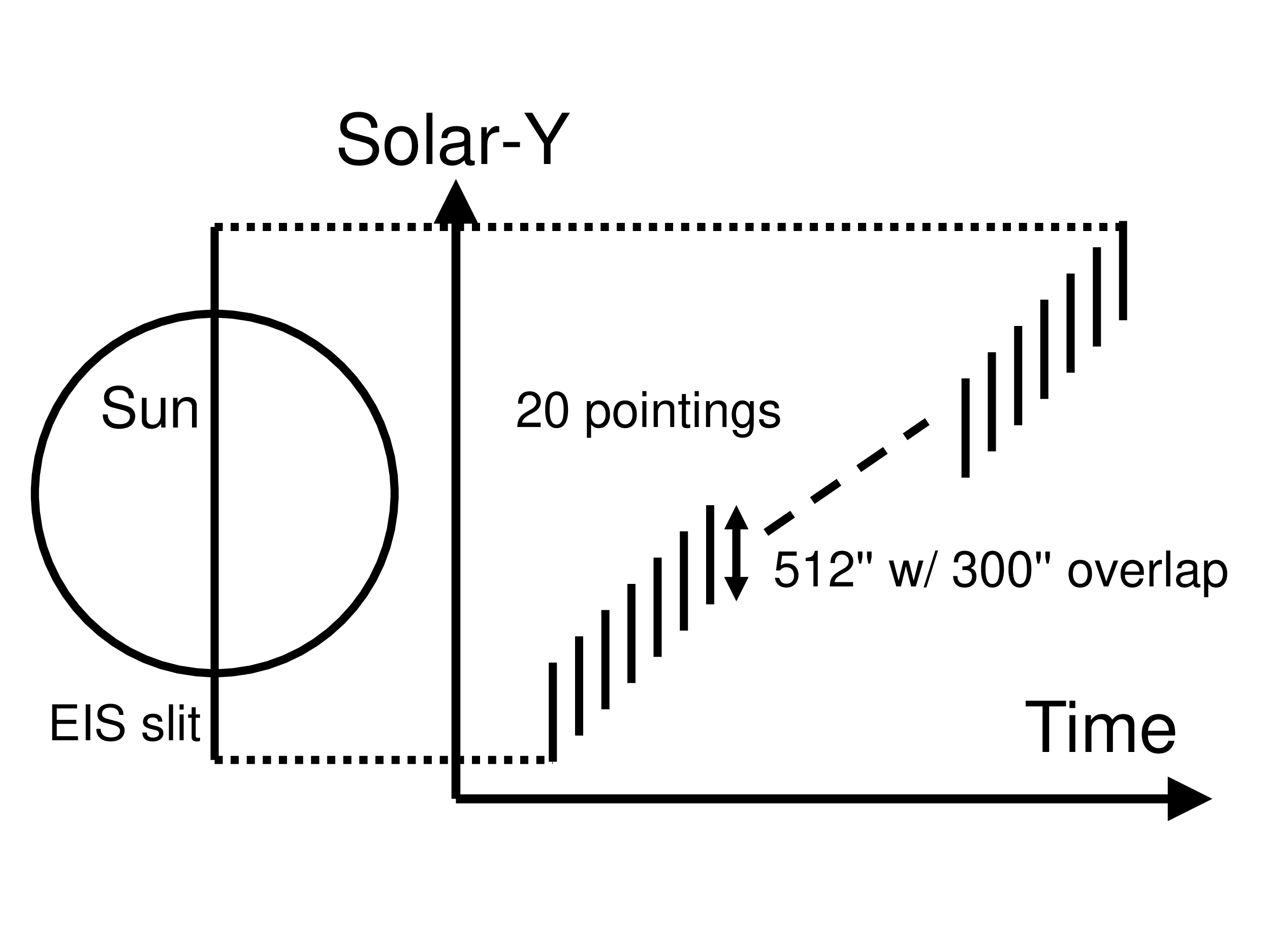}
  \caption{Schematic of north--south scanning in HOP79.}
  \label{fig:ns_scan_schem}
\end{figure}
We used the EIS spectral data of the north--south scans 
along the solar meridian acquired
by the \textit{Hinode} Observing Plan 79 (HOP79). 
Figure~\ref{fig:data_hop79} shows the context image 
with a {white} line indicating the location of EIS spectral data.
The pointing procedure in this observation
is schematized in Figure~\ref{fig:ns_scan_schem}.  
Using the $1'' \times 512''$ slit,
five exposures with $1''$ offset 
were conducted at each pointing. The slit is oriented
in the north-south direction, i.e., the solar-$y$ direction of the 
heliocentric coordinate. 
The pointing was shifted in the north-south direction with an overlap
of $300''$. In this way, we could investigate the center-to-limb variation of 
the Doppler shifts of the emission lines by mutual calibration 
among the overlapping scans.
The exposure time ($120 \, \mathrm{s}$) was sufficient  
for an appropriate signal-to-noise (S/N) ratio for many 
coronal emission lines even in the quiet region.  
The EIS study consisted of 16 spectral windows with 
widths of $24$--$48$ pixels ($\simeq 0.5$--$1.0${\AA}). 

We analysed the spectral data in October 
%\textcolor{red}{
7--8
%}
and December 
%\textcolor{red}{
2--3
%}
of 2010. 
During this period, the solar activity was relatively low 
and the observed area contained no large coronal hole, thus
enabling us to avoid the influence of these regions and to focus only on  
the quiet regions.
%\textcolor{red}{
Note that, although the data was carefully chosen to avoid 
brighter loops,
the analysed quiet region still has structures and complex
magnetic connectivity as seen in Fig.~\ref{fig:data_hop79}.
%}

%%%%%%%%%%%%%%%%%%%%%%%%%%%%%%%%%%%%%%%%%%%%%%%%%%%%%%%%%%%%%%%%%%%%%%%%

\section{Data reduction and analysis}
  \label{sect:cal_data_reduc}

Eleven of the observed emission lines were selected for a detailed 
analysis of their Doppler shifts (Table \ref{tab:qr_dop}).
The selection was made  by comparing the line profiles
at two locations; beyond the solar limb and on the disk.
When the emission-line intensity signal was significantly larger 
in the disk observation than in the area beyond the limb,
the emission line was selected for further analysis.
Emission lines with known blending effects from neighboring lines were also
removed from the candidates. The details of the 
selection process are presented in the Appendix.

\begin{figure}[!bp]
  \centering
  \includegraphics[width=8.4cm,clip]{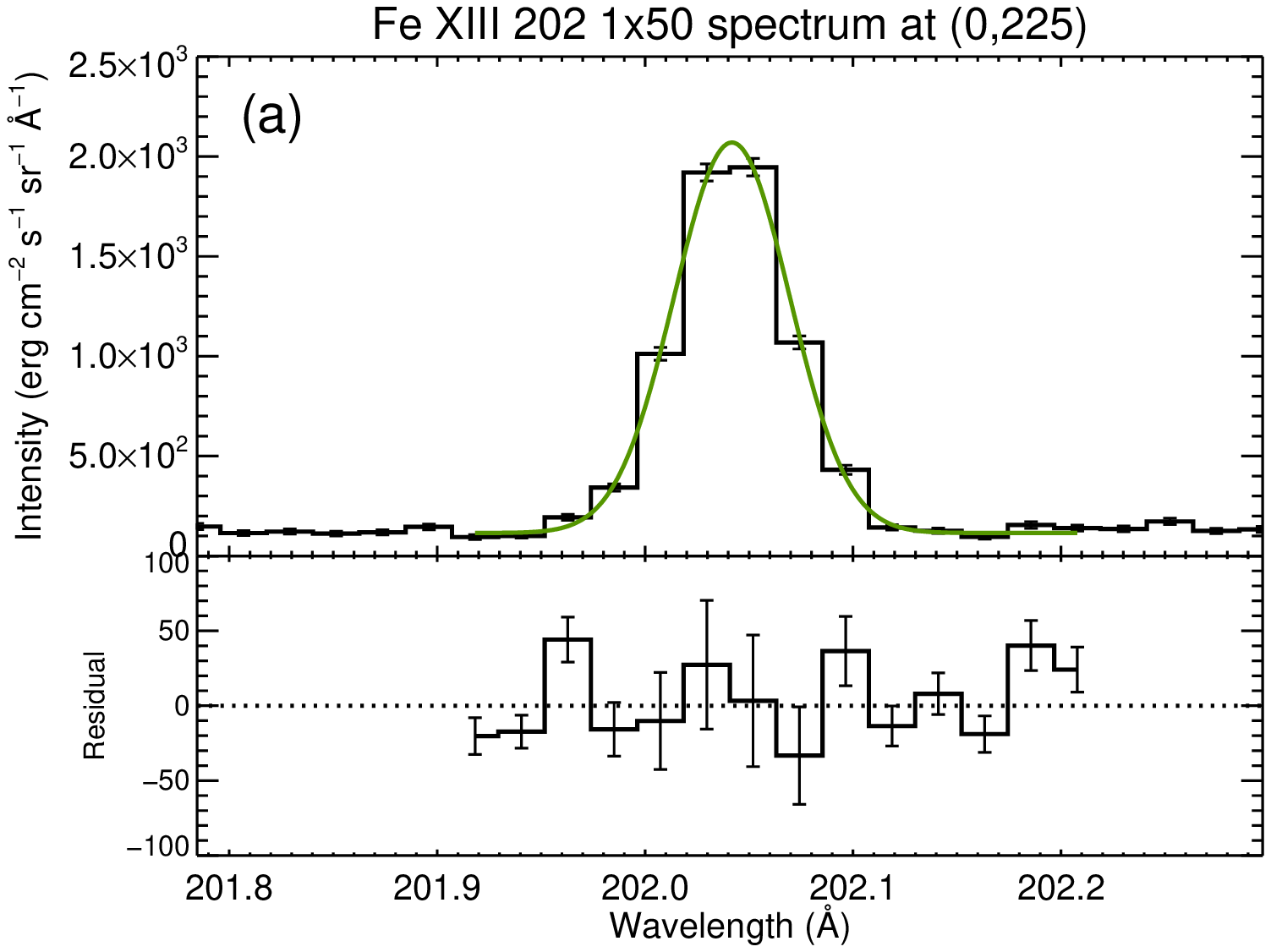}
  \includegraphics[width=8.4cm,clip]{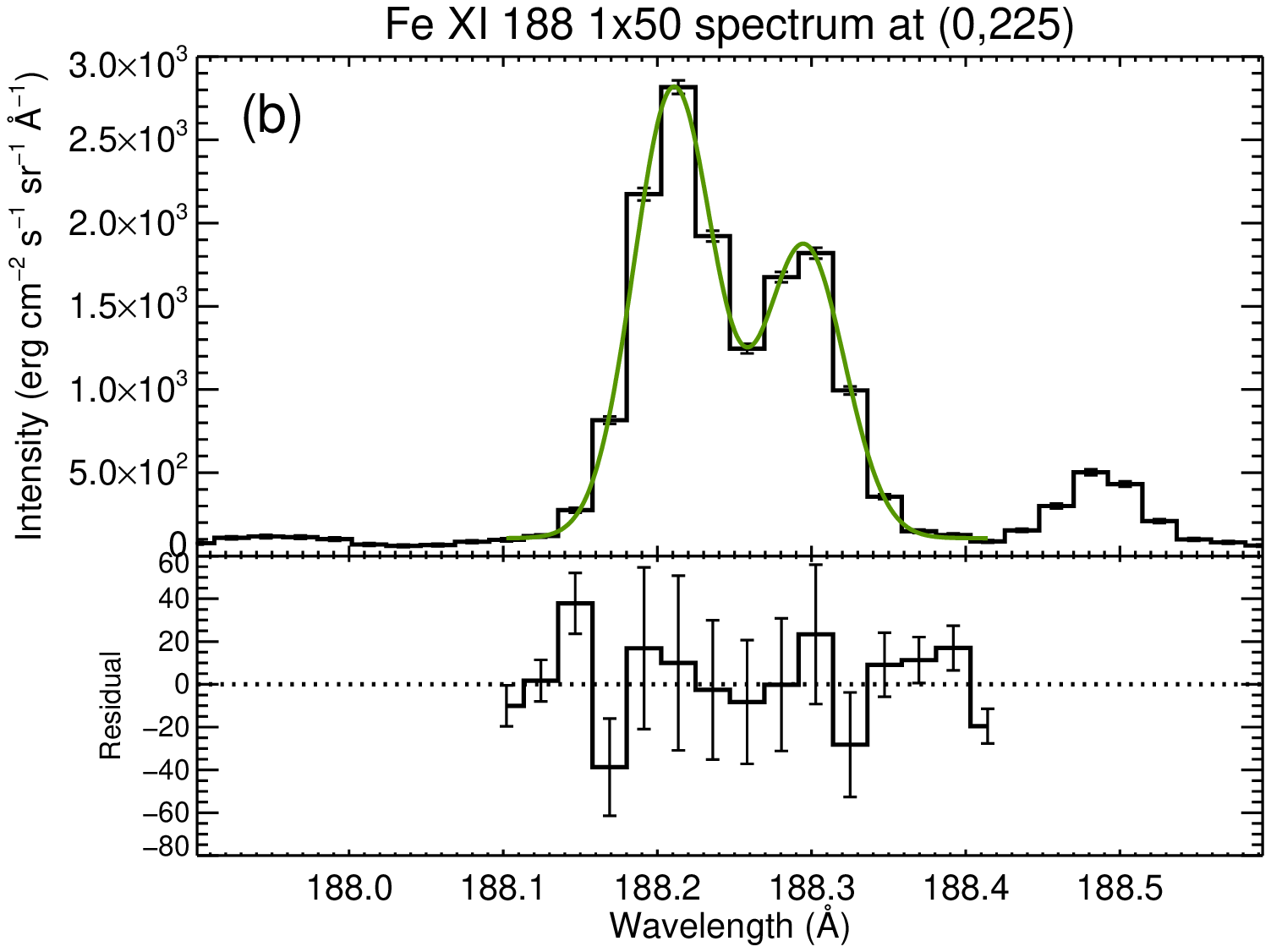}
  \caption{Examples of the observed emission lines and their fitting results.
(a) Fe \textsc{xiii} $202.04${\AA}  
and 
(b) Fe \textsc{xi} $188.21${\AA}/$188.30${\AA}
in the October observation sequence.
The observational data is in the histograms, and the fitted functions are
in solid green lines in the upper subpanels.
Lower subpanels show the fitting residuals with error bars over 
the fitting range.
The error bars include photon noise and uncertainty 
in the CCD pedestal and dark current.}
  \label{fig:lp_ex_dc}
\end{figure}

Each emission line was fitted to a Gaussian function.
Prior to the fitting, the spectra were spatially integrated over
$50''$ in the solar-$y$ direction to 
reduce the fluctuations introduced by 
coronal structures ({e.g.}, bright points) and non-radial motions.
An example is shown in Figure~\ref{fig:lp_ex_dc}(a).
The residuals are within $\sim 2${\%} of the peak in the spectrum 
and are comparable to the photon noise.  
Most of the eleven selected lines were fitted by a single-component 
Gaussian function.
The fitting range of the wavelength was $8$--$14$ pixels including the line
centroid. The Fe \textsc{xi} $188.21${\AA} and $188.30${\AA} 
lines, which overlap, were combined and simultaneously fitted 
by a double-component Gaussian function (Fig.~\ref{fig:lp_ex_dc}b). 

\begin{figure}[!bp]
  \centering
  \includegraphics[width=8.4cm]{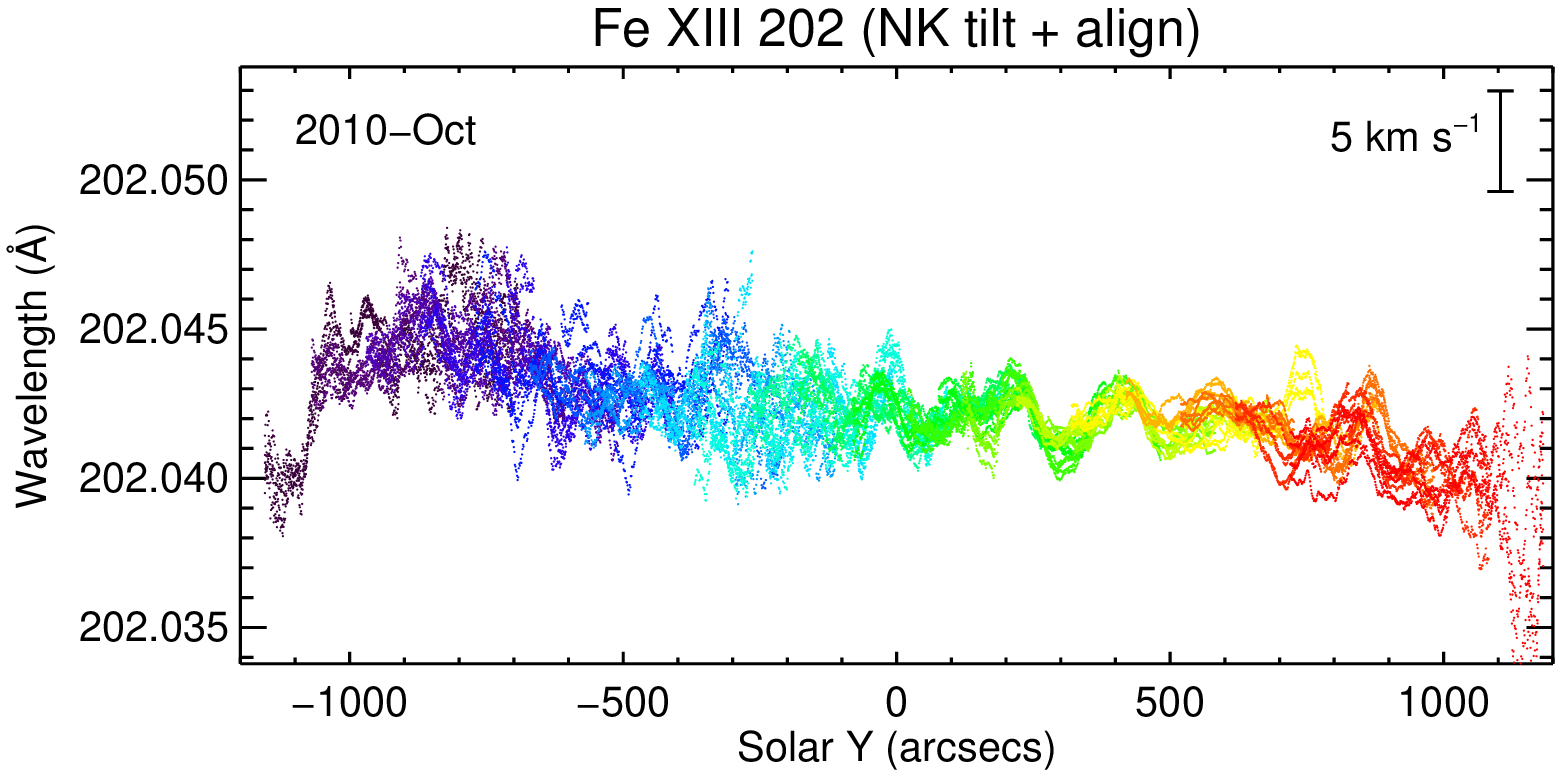}
  \includegraphics[width=8.4cm]{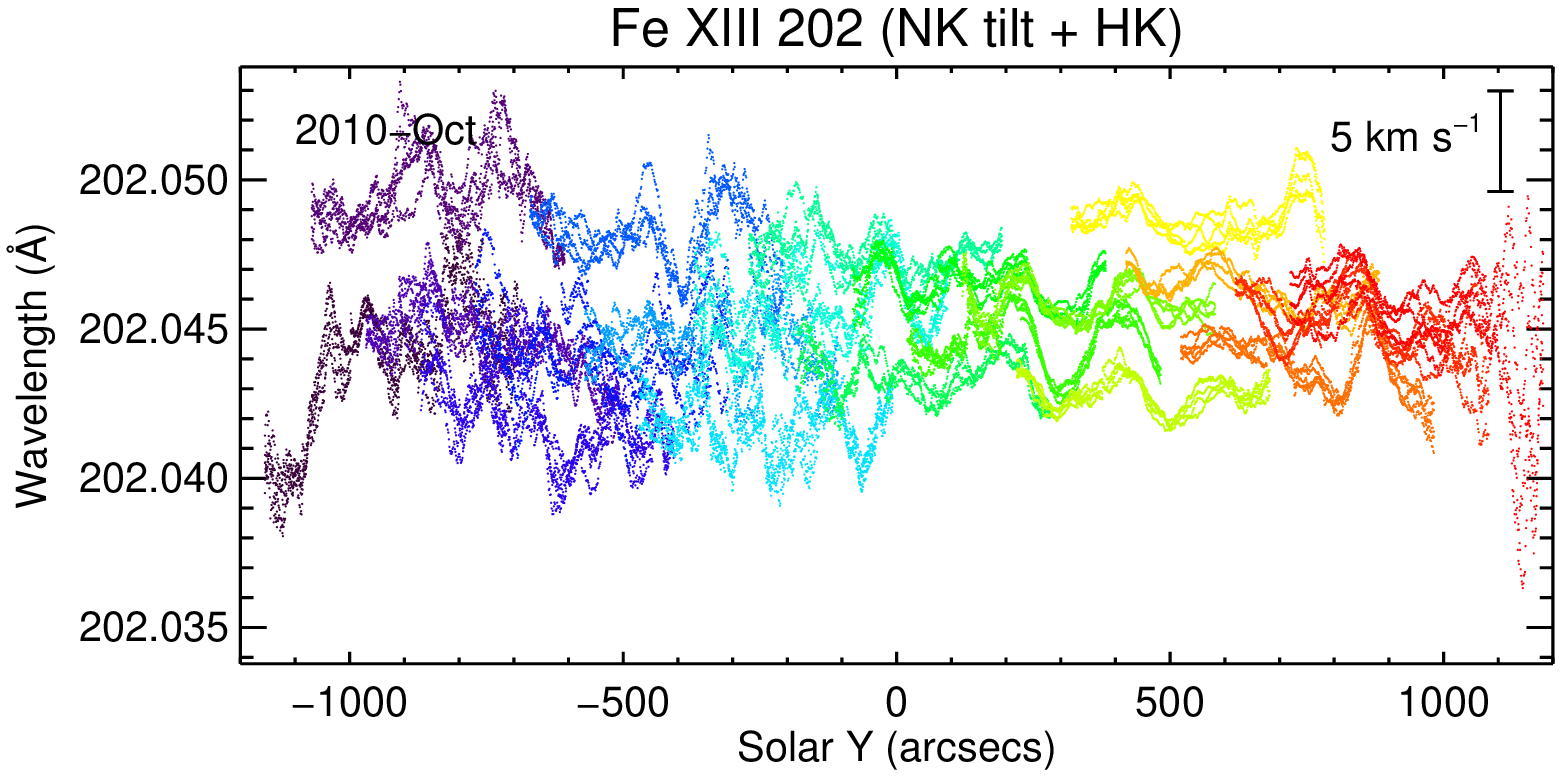}
  \caption{Line centroid wavelength as functions of solar-$y$.
Different colors indicate data collected at different north-south 
pointings ($20$ locations). Note that each data segment for each pointing
is represented by a unique color (i.e., each of the
20 colors represents one of the 20 segments) .
The apparent gradation behavior is attributed to overlapping segments.
{Top}: Data calibrated by our alignment procedure described in the text.  
{Bottom}: Data calibrated by the standard procedure provided by 
the SolarSoftware package.
}
  \label{fig:cal_align}
\end{figure}

The wavelengths of each exposure and pointing are relatively offset 
from each other.
We removed these offsets by the following procedures: 
(1) Five profiles in the exposures in each north-south pointing were aligned 
by changing their wavelength scales 
to remove their (albeit tiny) relative offsets, 
and (2) the aligned profiles for each pointing were further aligned 
with their neighboring ones using the profiles in 
the spatially overlapping observing regions.
An example of an alignment is shown in the top panel of 
Figure~\ref{fig:cal_align}. For comparison, the results of the
standard SolarSoftware package are presented in the bottom panel. 
After our alignment procedures, the line centroid wavelengths were 
consistent within $0.002${\AA} ($\simeq 3 \, \kmpers$),
thus demonstrating clear improvement over the standard procedure.

\begin{figure}[!bp]
  \centering
  \includegraphics[width=12.2cm,clip]{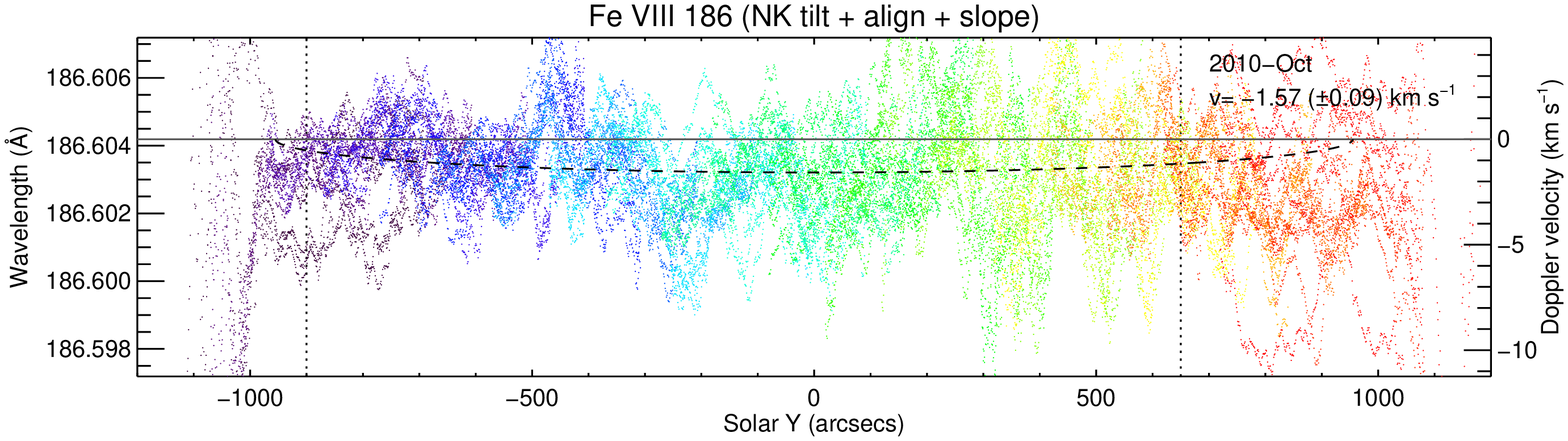}
  \includegraphics[width=12.2cm,clip]{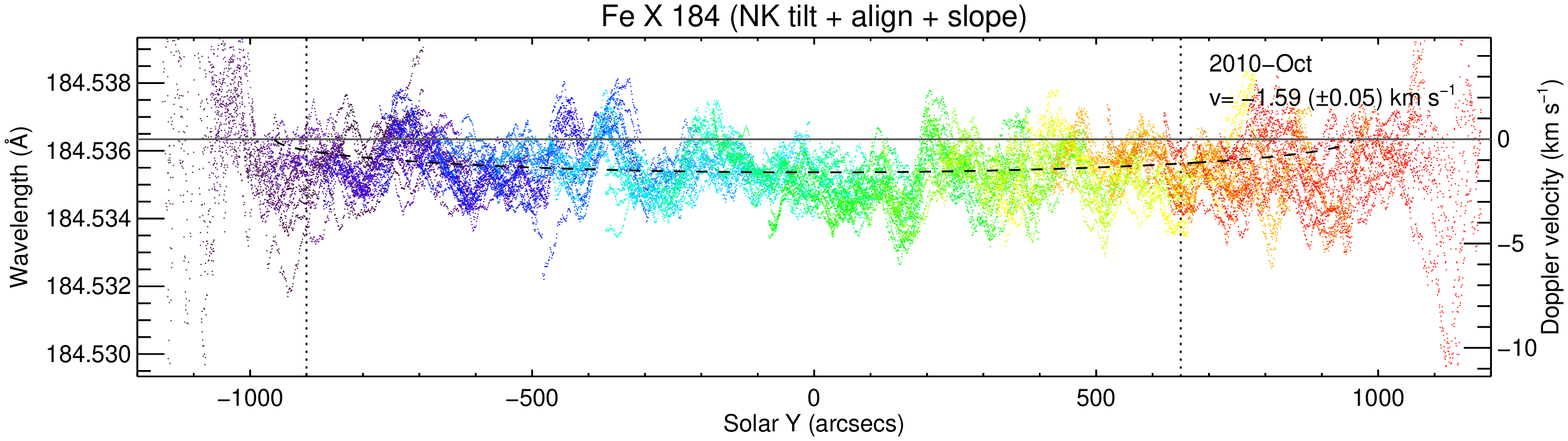}
  \includegraphics[width=12.2cm,clip]{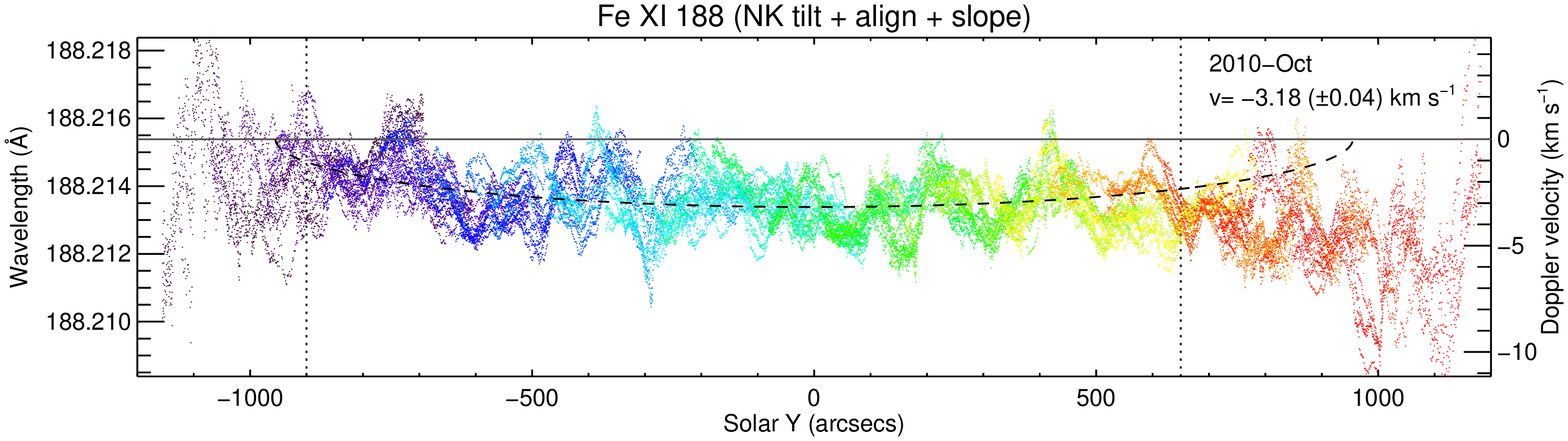}
  \includegraphics[width=12.2cm,clip]{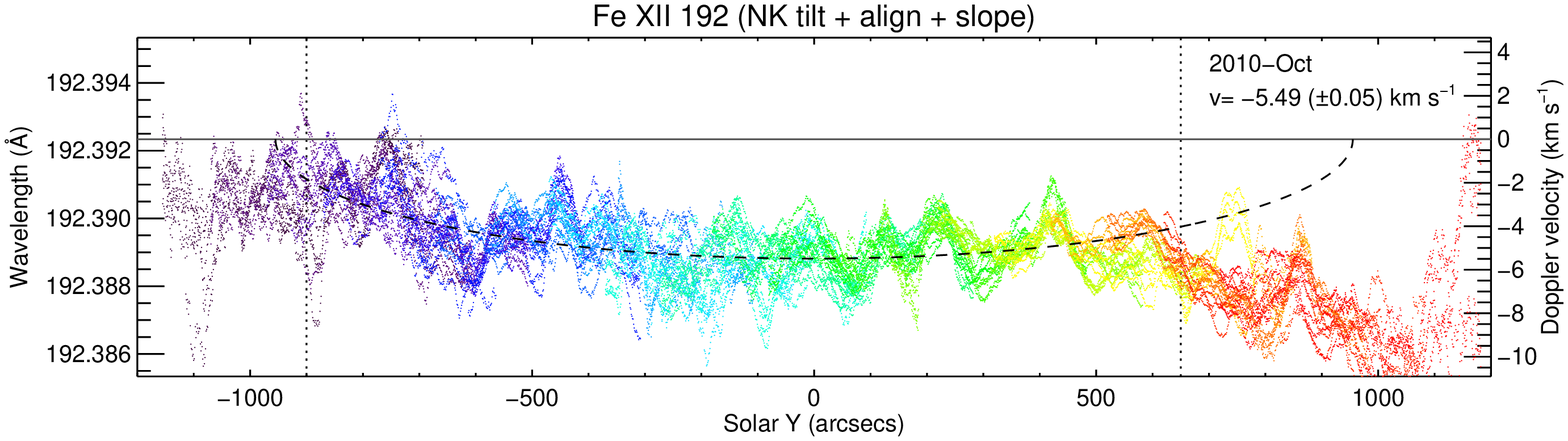}
  \includegraphics[width=12.2cm,clip]{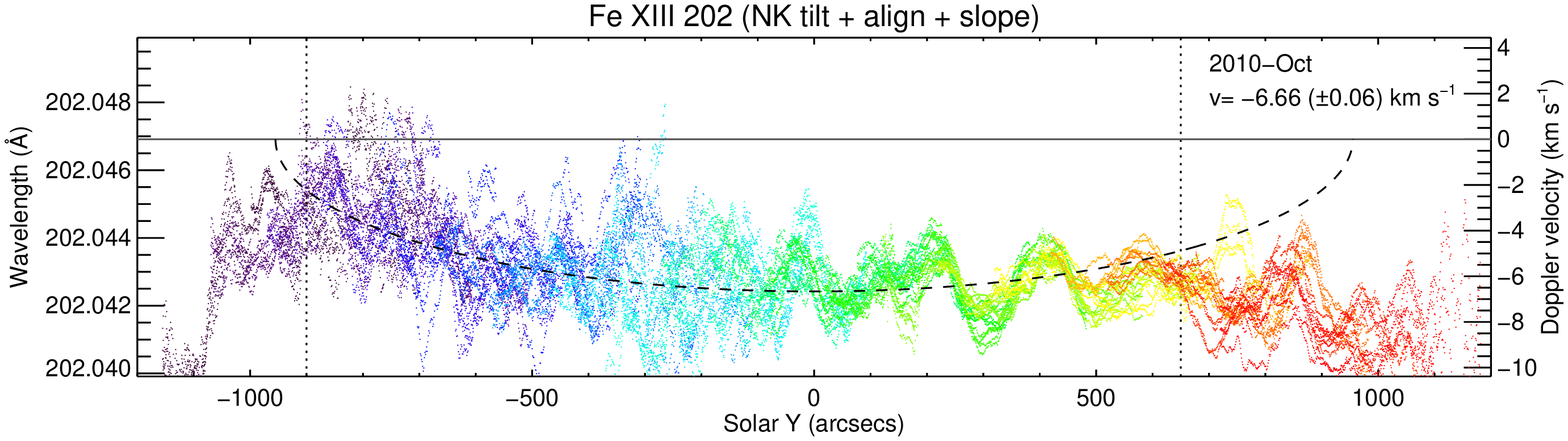}
  \caption{Centroid wavelengths of the emission lines as 
  functions of solar-$y$,
   i.e., the center-to-limb variation of the Doppler shift. Each panel
    represents a different emission line:
    (a) Fe \textsc{viii} $186.60${\AA},
    (b) Fe \textsc{x} $184.54${\AA},
    (c) Fe \textsc{xi} $188.21${\AA}, 
    (d) Fe \textsc{xii} $192.39${\AA}, and 
    (e) Fe \textsc{xiii} $202.04${\AA}.
    In each panel, the dashed line is the fitting curve based on 
    the radial flow model, and
    the vertical dotted lines indicate the fitting range.
  }
  \label{fig:limb2limb_var_Oct}
\end{figure}

Figure~\ref{fig:limb2limb_var_Oct} plots 
the Doppler shifts as functions of solar-$y$.
The final procedure calibrates the absolute velocity;
in other words, the zero-velocity wavelength determination. Here, the
spatial distribution of the Doppler shift was fitted by
a radial flow model of the form
$v(\theta) = v_0 \cos \theta$ where $v_0$ is the radial velocity and 
$\theta$ is the angle between the line of sight and the normal to the 
solar surface.  The solar-$y$ is represented as $y=R_{\odot} \sin \theta$.  
To fit the data, we converted the abscissa into $\cos\theta$ and
applied a linear function. One of the fitting parameters was the 
interception (i.e., wavelength shift) of the fitting line 
at the limb ($\cos \theta=0$).
The obtained finite interception was used to correct the entire
distribution assuming zero Doppler shift at the limb.
After this correction, the Doppler
shift at the disk center corresponded to the outflow velocity $v_0$.
The fitting was done within the range indicated 
by the region between the two vertical dotted lines in Fig. 
~\ref{fig:limb2limb_var_Oct}. 
The fitting error in $v_0$ was $\sigma_{\mathrm{fit}} < 0.1 \, \kmpers$.
Note that there is a small 
coronal hole at the north pole and the emission lines are clearly blueshifted 
at $y \geq 700''$, indicating possible outflow.  
The radial 
velocity at the disk center (stated in the upper-right corner of each panel in 
Fig.~\ref{fig:limb2limb_var_Oct}) reveals an
increasing magnitude of the shift 
({i.e.}, stronger upflows) as the line-formation temperature increases.

\begin{figure}[!bp]
  \centering
  \includegraphics[width=5.2cm,clip]{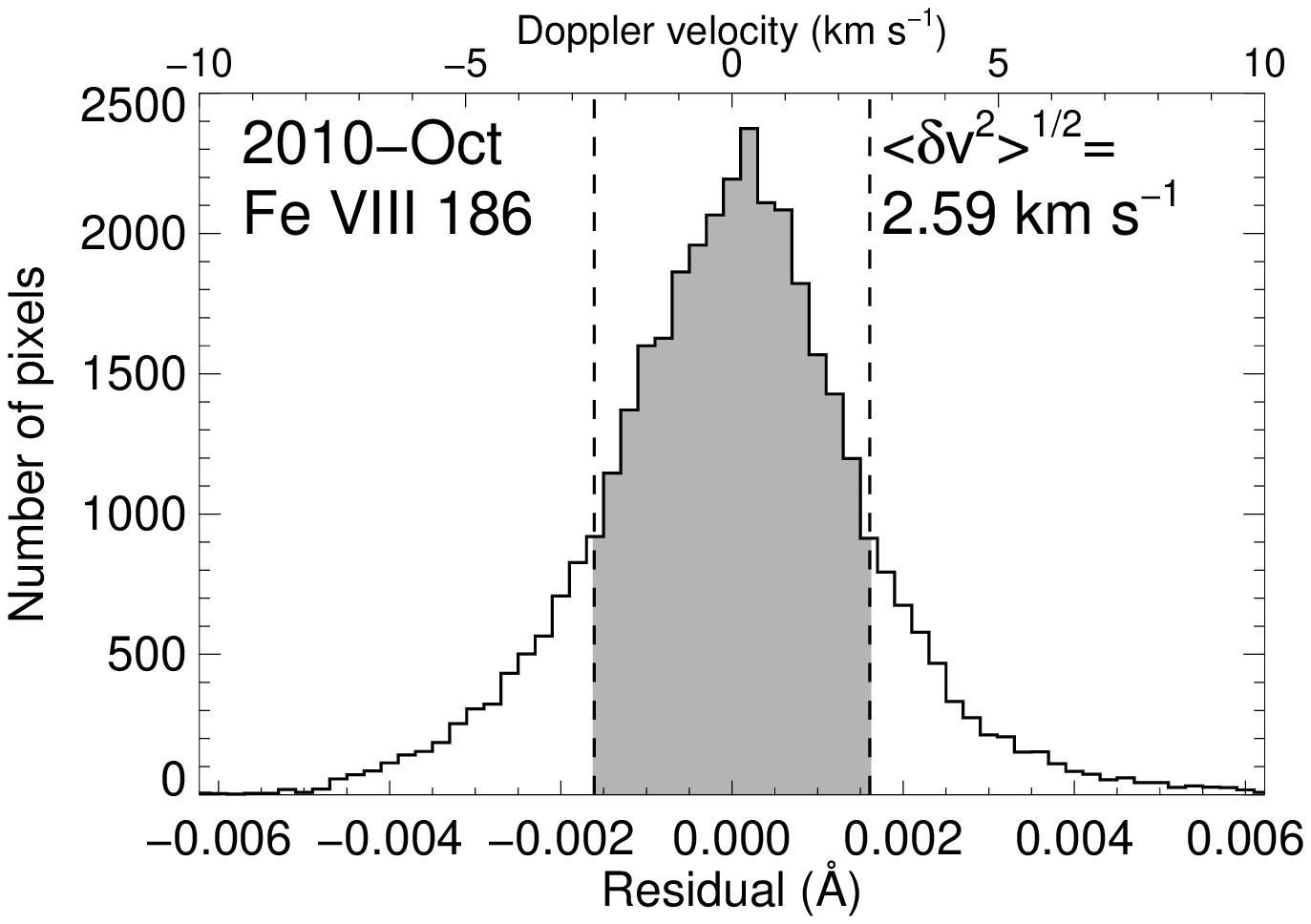}
  \includegraphics[width=5.2cm,clip]{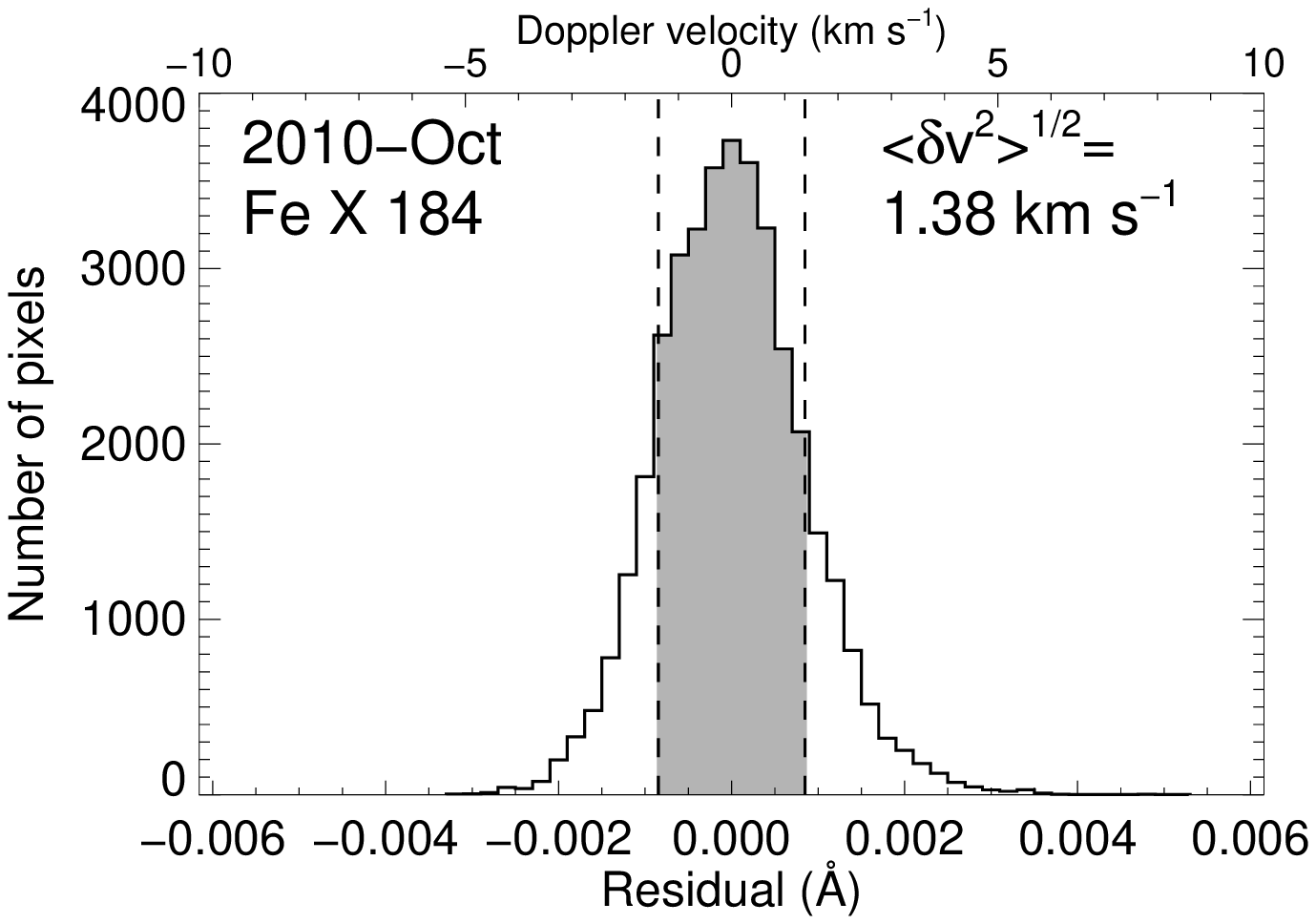}
  \includegraphics[width=5.2cm,clip]{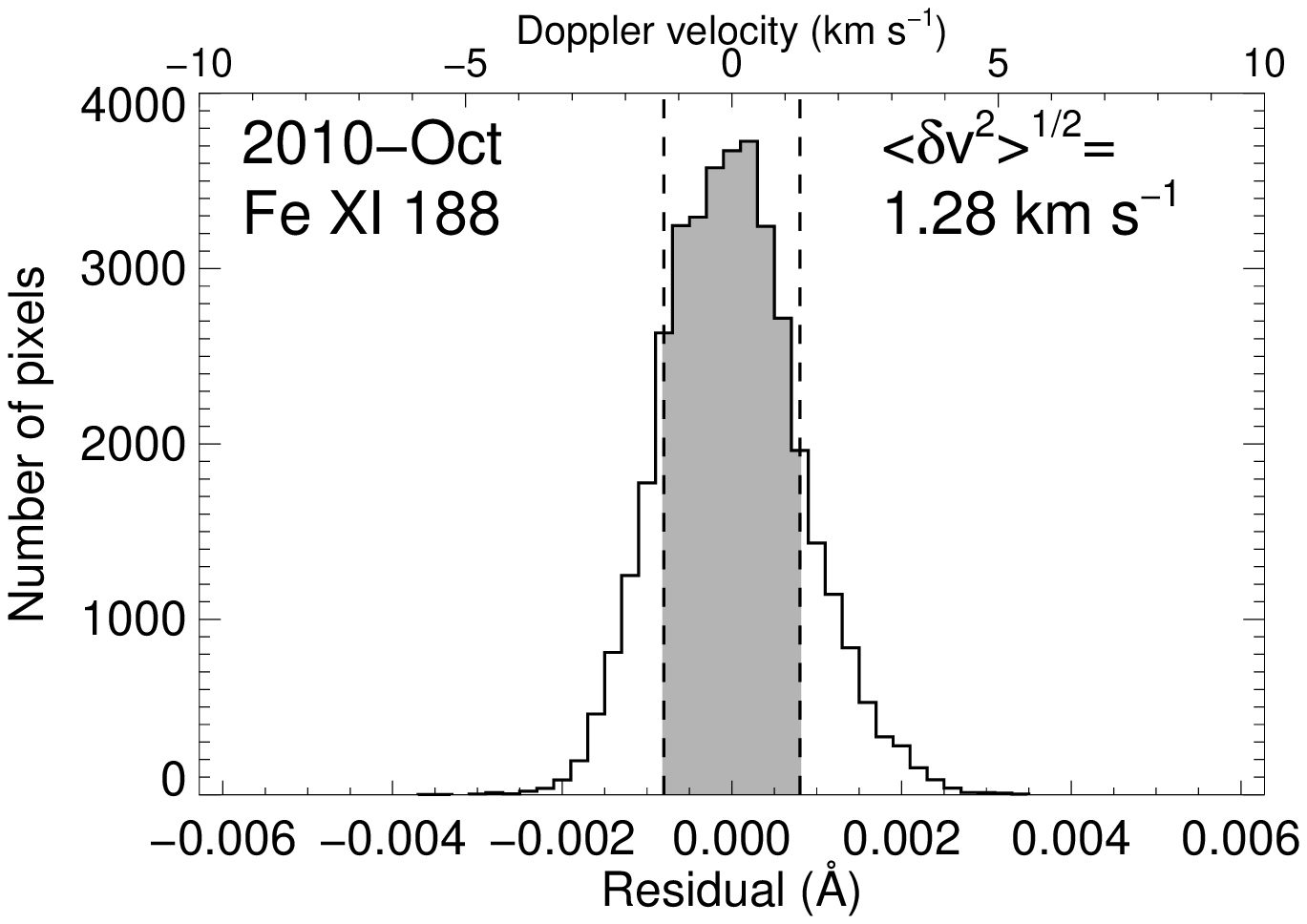}
  \includegraphics[width=5.2cm,clip]{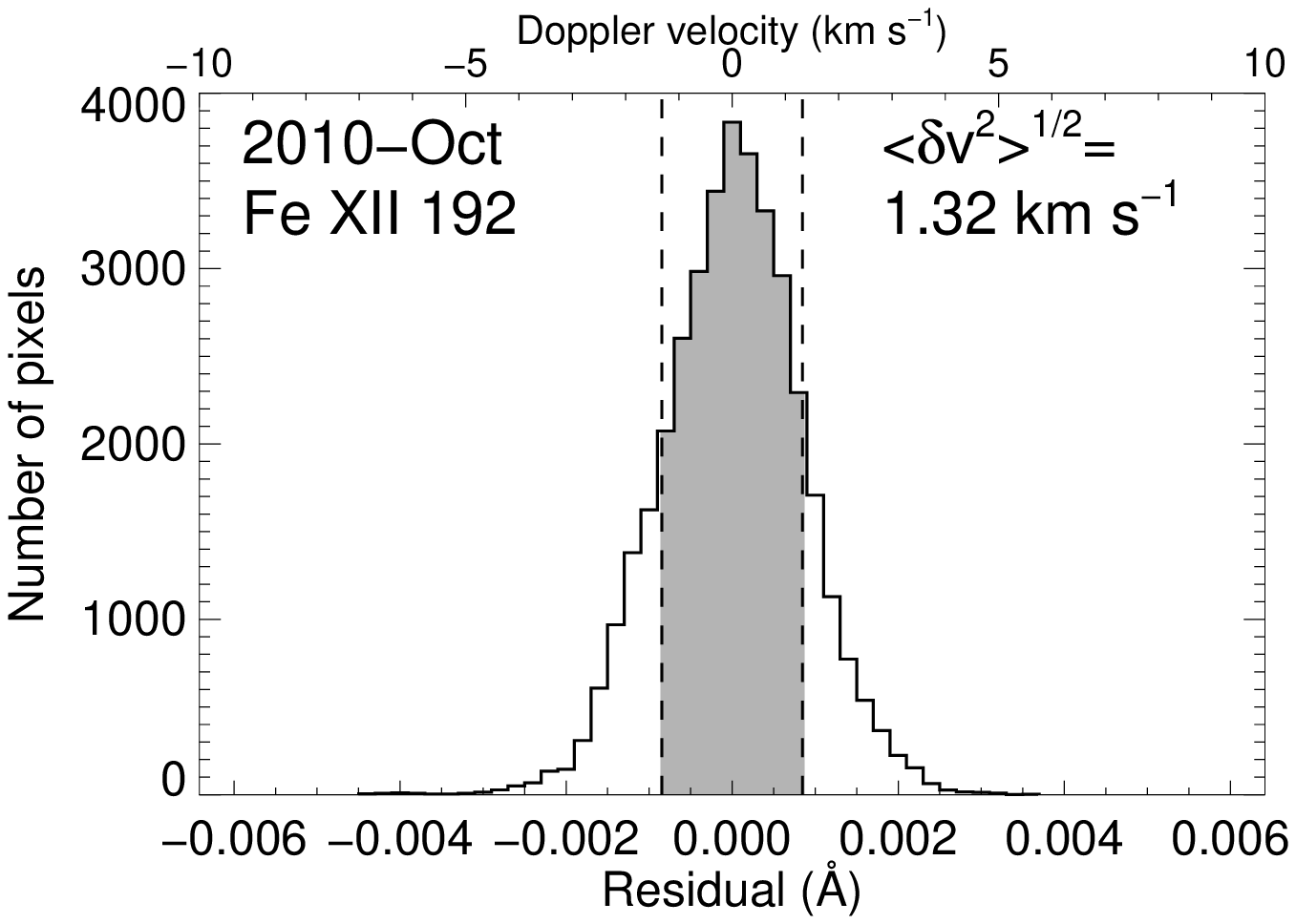}
  \includegraphics[width=5.2cm,clip]{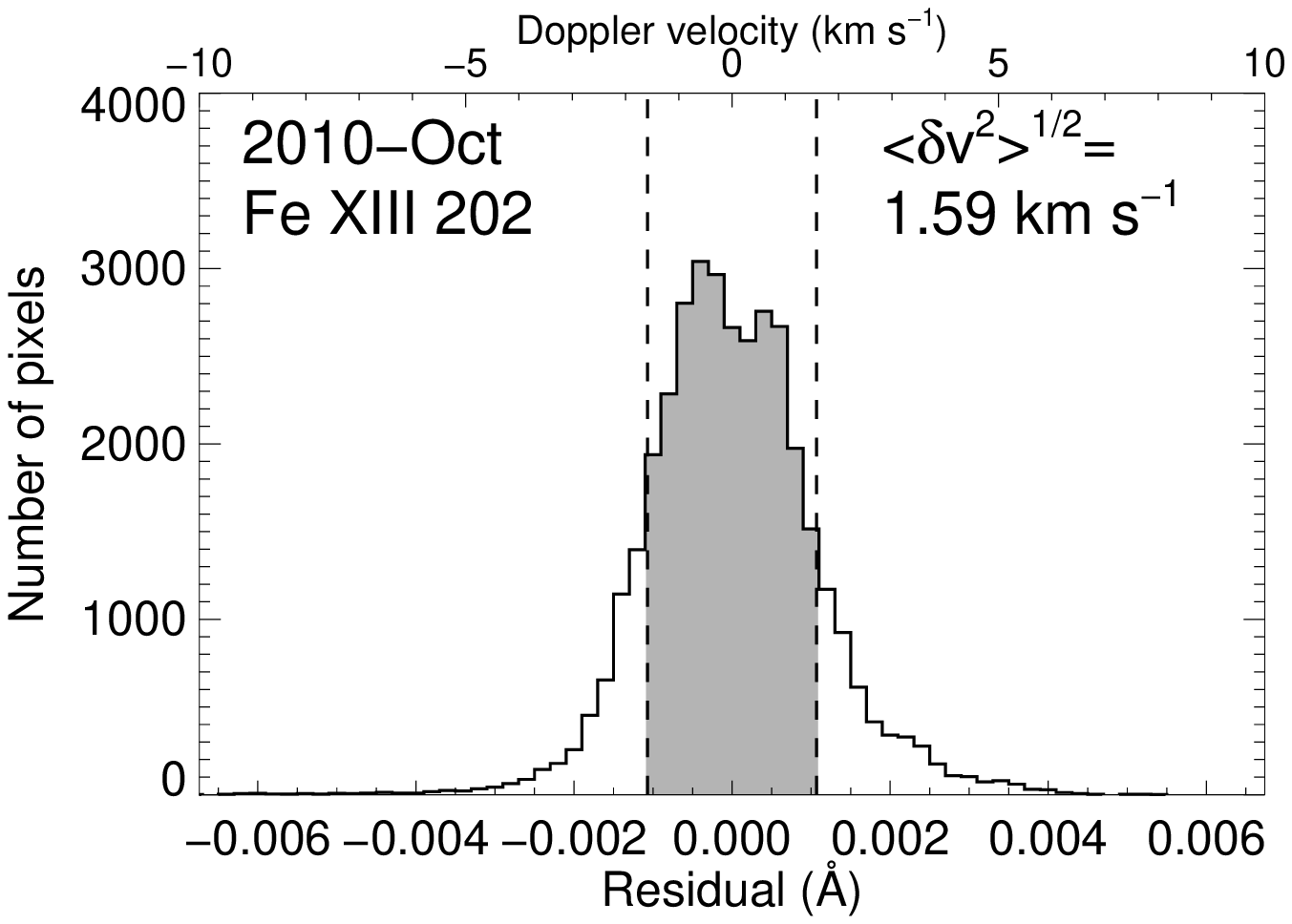}
  \caption{Histograms of deviation in
the actual Doppler shifts from the fittings based on the 
radial flow model (Fig.~\ref{fig:limb2limb_var_Oct}).  Results are plotted for 
the Fe \textsc{viii}, and \textsc{x}--\textsc{xiii} lines.  
Numbers in the upper-right corners are the standard deviations
$\langle \delta v^2 \rangle^{1/2}$, 
also shown by the vertical dashed lines.}
  \label{fig:cal_limb2limb_res}
\end{figure}
Figure~\ref{fig:cal_limb2limb_res} shows histograms of the deviation in the 
Doppler shift from the fitted radial flow model 
(Fig.~\ref{fig:limb2limb_var_Oct}).
The standard deviation 
was $\langle \delta v^2 \rangle^{1/2}= 1\text{--}3 \, \kmpers$.
This finite deviation $\langle \delta v^2 \rangle^{1/2}$, which should include 
the real flow fluctuations in the solar quiet regions,
affects our fitting results of the radial flow speed $v_0$.
Because we computed the \textit{averaged} flow speed 
beyond such spatial or temporal fluctuations,
we took $\sigma_{\mathrm{tot}} = 
\left( \sigma_{\mathrm{fit}}^{2} + \langle \delta v^2 \rangle 
\right)^{1/2}$ as the error in our estimated radial velocity $v_0$.

\section{Results}
  \label{sect:cal_results}

\begin{table}
  %\captionsetup{width=13cm}
  \centering
  \caption{Radial velocity $v_0$ ($\kmpers$) in our fitting model 
(interpretable as the Doppler velocity at the disk center).
$\sigma_{\mathrm{fit}}$ is the error in the fitted velocity and 
$\langle \delta v^2 \rangle^{1/2}$ is the standard deviation 
relative to the fitted curve.  
The symbol ${}^{\mathrm{b}}$ appended to the wavelength (denoted by 
Wvl.\ in the second column) means that the emission line is 
potentially blended with another emission line. } 
  \begin{tabular}{llrrrrrrr} %9 rows
    \toprule
    & 
    & 
    & \multicolumn{6}{c}{\small Radial velocities and their errors
      ($\mathrm{km} \, \mathrm{s}^{-1}$)} 
    \\
    \cmidrule(lr){4-9}
    &  
    & 
    & \multicolumn{3}{c}{\small October} 
    & \multicolumn{3}{c}{\small December} 
    \\
    \cmidrule(lr){4-6}
    \cmidrule(lr){7-9}
    \multicolumn{1}{l}{\footnotesize Ion}
    & \multicolumn{1}{l}{\footnotesize Wvl.~({\AA})} 
    & \multicolumn{1}{c}{\footnotesize $\log T \, [\mathrm{K}]$} 
    %\rule[-1pt]{0pt}{12pt}
    & \multicolumn{1}{c}{\footnotesize $v_0$}
    & \multicolumn{1}{c}{\footnotesize $\sigma_{\mathrm{fit}}$}
    & \multicolumn{1}{c}{\footnotesize $\langle \delta v^2 \rangle^{1/2}$}
    & \multicolumn{1}{c}{\footnotesize $v_0$}
    & \multicolumn{1}{c}{\footnotesize $\sigma_{\mathrm{fit}}$}
    & \multicolumn{1}{c}{\footnotesize $\langle \delta v^2 \rangle^{1/2}$} \\
    \midrule
    Fe \textsc{viii} & $186.60$ & $5.69$ & $-1.57$ & $0.09$ & $2.59$ & $-0.35$ & $0.13$ & $2.85$ \\
    Si \textsc{vii}  & $275.35$ & $5.80$ & $-2.45$ & $0.10$ & $2.86$ & $ 4.56$ & $0.16$ & $3.41$ \\
    Fe \textsc{ix}   & $188.49$ & $5.92$ & $ 1.00$ & $0.06$ & $1.73$ & $ 1.55$ & $0.09$ & $1.92$ \\
    Fe \textsc{x}    & $184.54$ & $6.04$ & $-1.59$ & $0.05$ & $1.38$ & $-1.84$ & $0.06$ & $1.37$ \\
                     & $257.26$ &        & $-3.63$ & $0.08$ & $2.05$ & $-2.09$ & $0.09$ & $1.91$ \\
    Fe \textsc{xi}   
    & $180.40^{\mathrm{b}}$ & $6.12$ & $-2.24$ & $0.07$ & $2.07$ & $ 0.48$ & $0.09$ & $2.00$ \\
                     & $188.21$ &        & $-3.18$ & $0.04$ & $1.28$ & $-1.12$ & $0.06$ & $1.27$ \\
                     & $188.30$ &        & $-3.26$ & $0.06$ & $1.34$ & $-0.44$ & $0.06$ & $1.31$ \\
    Fe \textsc{xii}  & $192.39$ & $6.19$ & $-5.49$ & $0.05$ & $1.32$ & $-3.18$ & $0.06$ & $1.28$ \\
    & $195.12^{\mathrm{b}}$ &        & $-2.38$ & $0.05$ & $1.29$ & $-0.05$ & $0.07$ & $1.44$ \\
    Fe \textsc{xiii} & $202.04$ & $6.25$ & $-6.66$ & $0.06$ & $1.59$ & $-6.57$ & $0.09$ & $1.91$ \\
    \bottomrule
  \end{tabular}
  \label{tab:qr_dop}
\end{table}
\begin{figure}[!bp]
  \centering
  \includegraphics[width=17.0cm,clip]{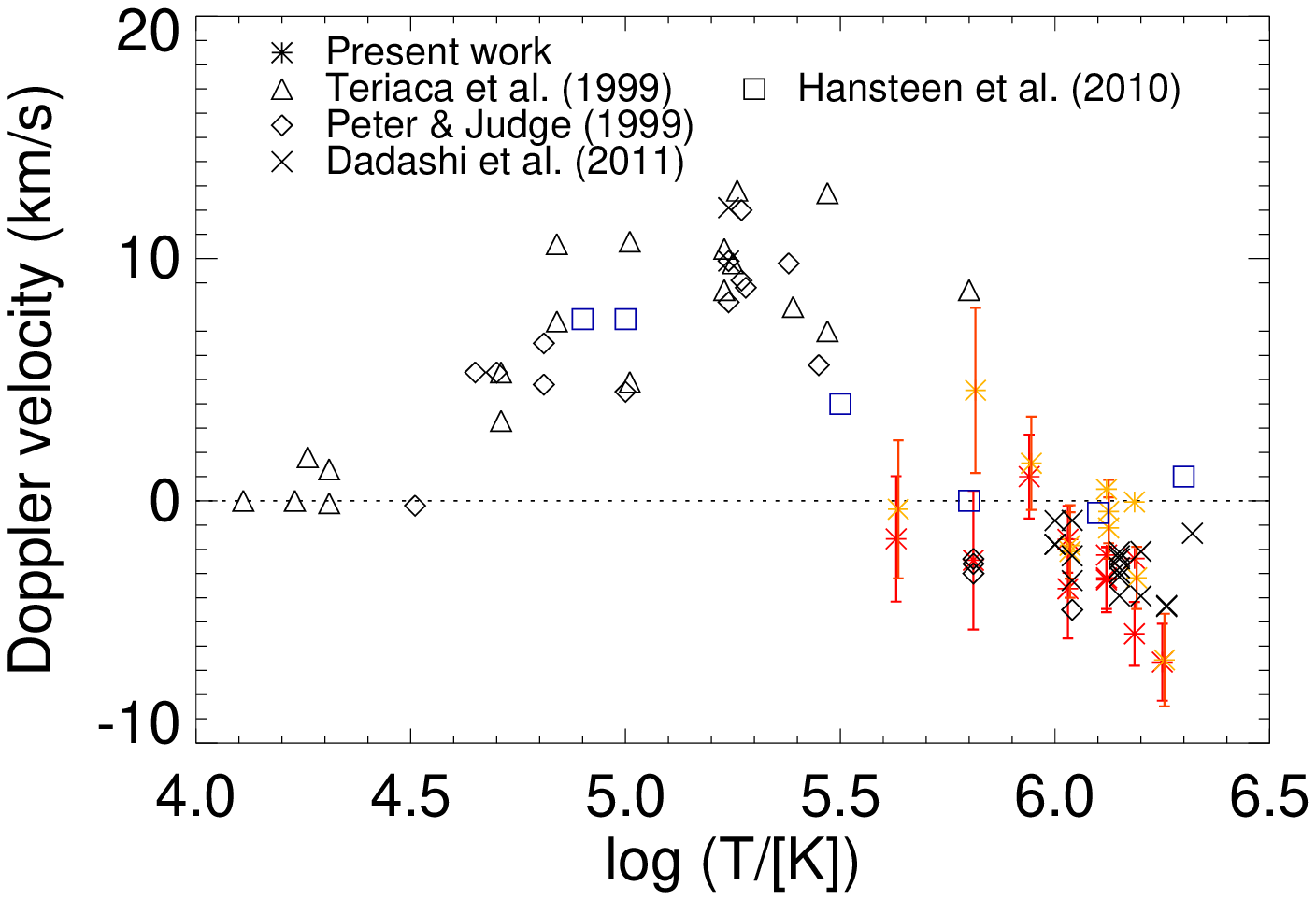}
  \caption{
Radial velocity $v_0$ 
(interpretable as the Doppler shift at the disk center) as a function 
of the line formation temperature.  
%\textcolor{red}{
Positive values indicate
red shifts (downflows) while negative values indicate blue shifts (upflows).
Red (orange) asterisks show the results in this study by
the October (December) data.  Astrerisk symbols without error bars show the 
results for the potentially blended emission lines Fe \textsc{xi} 
$180.40${\AA} and Fe \textsc{xii} $195.12${\AA}.  
The data points by previous observations are shown by triangles
\citep{teriaca1999}, diamonds \citep{peterjudge1999} and crosses
\citep{2011A&A...534A..90D}.
Blue diamonds are the data points synthesized by using the MHD
simulation results by \cite{hansteen2010}.
%\sout{
%Red squares (green diamonds)
%represent the October (December) data.  {Solid open symbols} show the
%results for the potentially blended emission lines Fe \textsc{xi}
%$180.40${\AA} and Fe \textsc{xii} $195.12${\AA}.
%The {Solid line} is the third-order polynomial function fitted
%to all data points excluding Fe \textsc{xi} $180.40${\AA} and
%Fe \textsc{xii} $195.12${\AA}.  The {gray} region between the two
%{dashed lines} indicates the standard deviation of the data relative to
%the fitted curve.}
%}
}
  \label{fig:qr_dop_cmpl}
\end{figure}
Obtained radial velocities $v_0$ from eleven emission lines for the 
temperature range $\logt = 5.7$--$6.3$ are listed in Table 
\ref{tab:qr_dop} and these results were plotted in 
Figure~\ref{fig:qr_dop_cmpl}.  
%\textcolor{red}{
%\sout{
%The solid line indicates a third order polynomial function fitted to the 
%all data points except for Fe \textsc{xi} $180.40${\AA} 
%and Fe \textsc{xii} $195.12${\AA}.  }}
The important conclusion here is that the Doppler shifts are almost 
zero or slightly positive ({i.e.,} downward) at the 
temperature below $\logt = 6.0$, and above that temperature the 
emission lines are blueshifted with increasing temperature, and the Doppler 
shift reaches $({-7})$ -- $({-6}) \, \kmpers$ at $\logt = 6.25$ 
(Fe \textsc{xiii}).

\section{Discussion}

To understand the dynamics in the solar transition region and lower corona,
we investigated the
Doppler shifts of quiet regions over a wide coronal temperature range
($\logt=5.7$--$6.3$) using the spectroscopic data taken
by the  \textit{Hinode} EIS.
By analyzing the data covering the meridian
from the south to the north poles, we successfully measure the shift
to an accuracy of $3 \, \kmpers$.  

Below $\logt = 6.0$ the Doppler 
shifts are almost zero or slightly positive ({i.e.,} 
downward); above this temperature, 
the Doppler shifts clearly become negative, reaching up to 
$-6 \, \kmpers$
%\textcolor{red}{
(see Fig.~\ref{fig:qr_dop_cmpl})
%}
.
In previous observation of Ne \textsc{viii} $770.43${\AA} 
in the quiet region at $\logt = 5.8$,
the Doppler shift was measured 
as $-2.6 \pm 2.2 \, \kmpers$ 
\citep{peter1999}, $-2.4 \pm 1.5 \, \kmpers$ \citep{peterjudge1999}, 
and $-1.9 \pm 2.0 \, \kmpers$ \citep{teriaca1999}.  Our results were in 
good agreement with those studies within the error margin.

\cite{2011A&A...534A..90D} measured
the Doppler shift of spectral lines formed between 1 and 2 MK
(
%\textcolor{red}{
$\logt=6.0$ and 6.3,
%}
 the coronal temperature range) in the quiet region
%\textcolor{red}{(
see Fig.~\ref{fig:qr_dop_cmpl})
%}
.
They combined the SUMER/\textit{SoHO} and
EIS observations around the disk center and established 
the absolute value at the reference temperature (1 MK)
by simultaneous wavelength calibration of the SUMER spectra.
They determined the velocity as
$(-1.8 \pm 0.6)\,\kmpers$ at 1 MK, peaking at
$(-4.4 \pm 2.2)\,\kmpers$ around 1.8 MK ($\logt=6.25$) , 
and dropping to $(-1.3 \pm 2.6)\,\kmpers$
at 2.1 MK ($\logt=6.32$).
Our results are consistent with theirs. 
Despite being based on different calibration methods,
our study and that of \cite{2011A&A...534A..90D}, commonly found
a blueshift of $(-6)$--$(-4)\,\kmpers$, indicating a
substantial upflow in the lower corona.

To explain the redshifts in the transition regions,
\cite{1982ApJ...255..743A} and \cite{roussel-dupre1982} conjectured
the return of previously heated spicule material.
\cite{hansteen1993} argued that redshifts arise from 
waves generated by nanoflares in the corona.
\cite{2006ApJ...638.1086P} reproduced
the redshifts in their three-dimensional 
simulations although they could not clarify the mechanisms.
\cite{2011A&A...531A..97Z} proposed that these downflows are
signatures of cooling plasma that drains from the coronal volume.
Based on 
their comprehensive study using three-dimensional 
magnetohydrodynamic (MHD) models,
\cite{hansteen2010} interpreted that
the pervasive redshifts in the transition regions 
and the weak upflows in the low corona 
are caused by rapid episodic heating at low heights of the upper chromosphere
to coronal temperatures producing downflows in the transition region lines.
Our observed near-zero Doppler shifts in the emission lines $\logt = 6.0$ 
are consistent with 
the synthesized Doppler shifts in \cite{hansteen2010}
%\textcolor{red}{
(see Fig.~\ref{fig:qr_dop_cmpl})
%}
.
On the other hand, the increasing magnitude of the
blueshift at higher temperature reaching 
$- 6.3 \pm 2.1 \, \kmpers$ at $\logt=6.25$ is substantially larger
than that predicted by the MHD model. This discrepancy of larger blueshifts 
was also observed by \cite{2011A&A...534A..90D}, suggesting that
the contributing solar heating events 
are more episodic and stronger than those
presumed in Hansteen et al.'s MHD models.

For further understanding,
we could investigate spatially and temporally resolved spectra.
\cite{2013A&A...557A.126W} studied the Doppler shift 
in the network and internetwork regions and found enhancements 
of the Doppler shift
magnitude and the non-thermal line width in the network regions 
than those in the internetwork regions. However, the physical interpretation
of this difference remains unclear.
In our measurements, the Doppler shifts similarly fluctuated on 
a scale of $100''$ by a few $\kmpers$. Since this spatial scale
is similar but a few times larger than that of the chromospheric network, 
we cannot attribute this fluctuation to 
the network structure studied by \cite{2013A&A...557A.126W}.
Therefore, we must compare not only the detailed 
structures in the transition region and the corona but also
the magnetic structures in the chromosphere.
%\textcolor{red}{
As stated above, previous studies and our results suggest 
that relatively short lived heating followed by cooling may play a major role
\citep[e.g.,][]{hansteen2010}. An effort should be made to 
look at Doppler shifts as a function of time over individual locations. 
Clearly, there are wavelength calibration issues associated with this, 
but if the situation is dynamic, analyzing seriously averaged data 
such as is done in this study may have reached the limit of its usefulness.
%}

Our calibration procedure is modeled on the
center-to-limb variation in the Doppler shift assuming that 
radial flows are uniformly distributed in the atmosphere at 
each temperature. This method, originally proposed by \cite{peter1999} 
and \cite{peterjudge1999}, was here improved by exploiting the
spectral data in the overlapping FOV. 
As shown in Fig. \ref{fig:cal_align} of Section \ref{sect:cal_data_reduc},
when the standard analysis package
in SolarSoftware was applied to the EIS data, the residual error exceeded
$5\,\kmpers$. By devising a sequential connecting procedure among
neighboring pointing data, we improved the precision to
better than $3\,\kmpers$.

Our method, although very simple,
allows flexible usage of the data because it requires no
reference spectrum. That is, we can analyze the
EIS data \textit{without} requiring simultaneous
SUMER observations.
%\textcolor{red}{
%\sout{On the downside,} 
On the other hand,
%} 
our procedure demands a relatively large field of view
covering the limb toward as further as the center. 
To improve the utility and reliability of our
method, we need to further
study its robustness. The zero-velocity assumption 
at and beyond the limb needs to be confirmed or limited.
To quantify the fluctuation in the Doppler shift along the limb, we could
analyze the observed spectral data
covering the limb circle.
Moreover, the applicability of the
center-to-limb calibration method should be generalized from covering only the
north-south meridian to coverage at any
latitude. For this purpose, we must study 
the solar rotation in the transition region or the corona.

The results obtained in this paper will provide a reference for the 
Doppler shifts of outflow in active regions
\citep{2015ApJ...805...97K,2014arXiv1411.4742K,kitagawa2010,2012ApJ...759...15M,hara2008}.

\acknowledgments

\textit{Hinode} is a Japanese mission developed and launched by
ISAS/JAXA, collaborating with NAOJ as a domestic partner,
NASA and STFC (UK) as international partners. Scientific
operation of the \textit{Hinode} mission is conducted by the \textit{Hinode}
science team organized at ISAS/JAXA. This team mainly
consists of scientists from institutes in the partner countries.
Support for the post-launch operation is provided by JAXA and
NAOJ (Japan), STFC (UK), NASA, ESA, and NSC (Norway).
 This research is supported by JSPS KAKENHI Grant Number 15H03640.

\appendix

\section{Selection of analysis lines}

The emission lines analysed in this study were selected
by investigating their observed profiles.  
%\textcolor{red}{
In this Appendix, we describe why some emission lines are not
selected for the analysis by showing each profile
even though the data itself was acquired. 
%}
Figure~\ref{fig:cal_lp_ex} shows the line profiles 
on the solar disk ($y=-750''$; {solid line}) and above 
the limb ($y=-1050''$; {dashed line}) in all EIS spectral windows 
of HOP79 observations taken on 2010 October 7--8.
The line profiles were integrated and averaged over
$100''$ spans in the solar-$y$ direction.
%\textcolor{red}{
Note that, this wider range ($100''$) than that for the Doppler analysis
in the main part of this paper is chosen only for
demonstrating that the rejected lines are already weak and noisy
even after averaging over wider area so that they are not useful
for the Doppler analysis. 
%}

\begin{figure}
  \centering
  \includegraphics[width=7.0cm,clip]{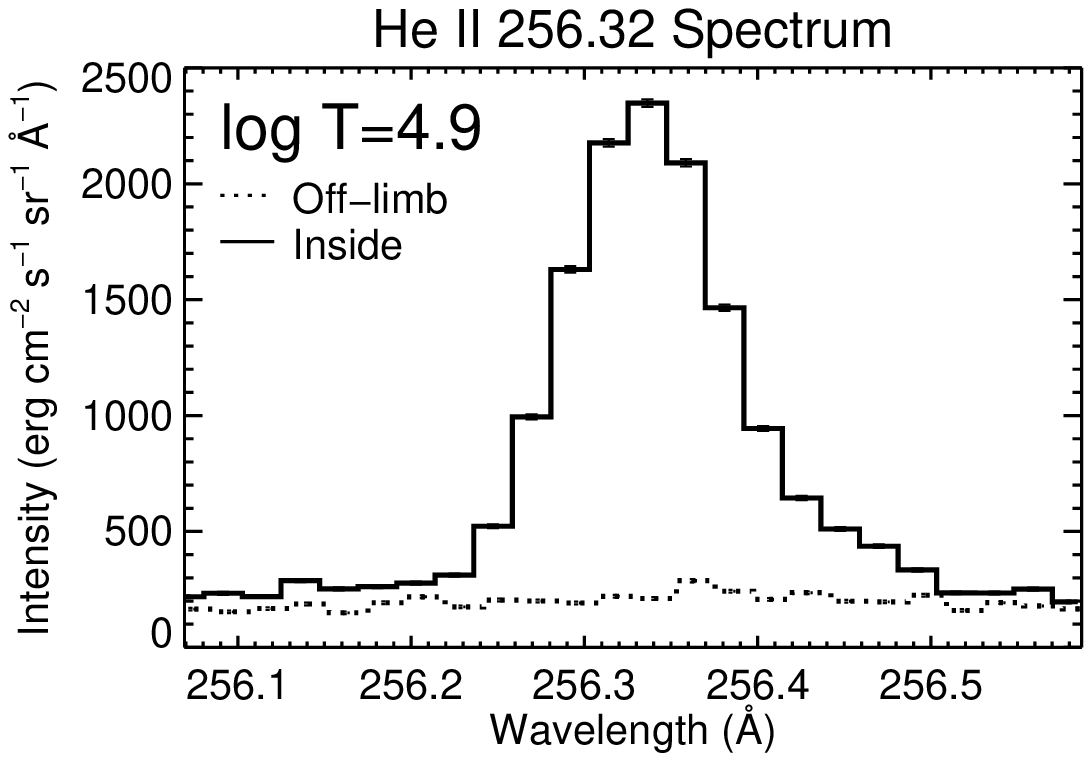}
  \includegraphics[width=7.0cm,clip]{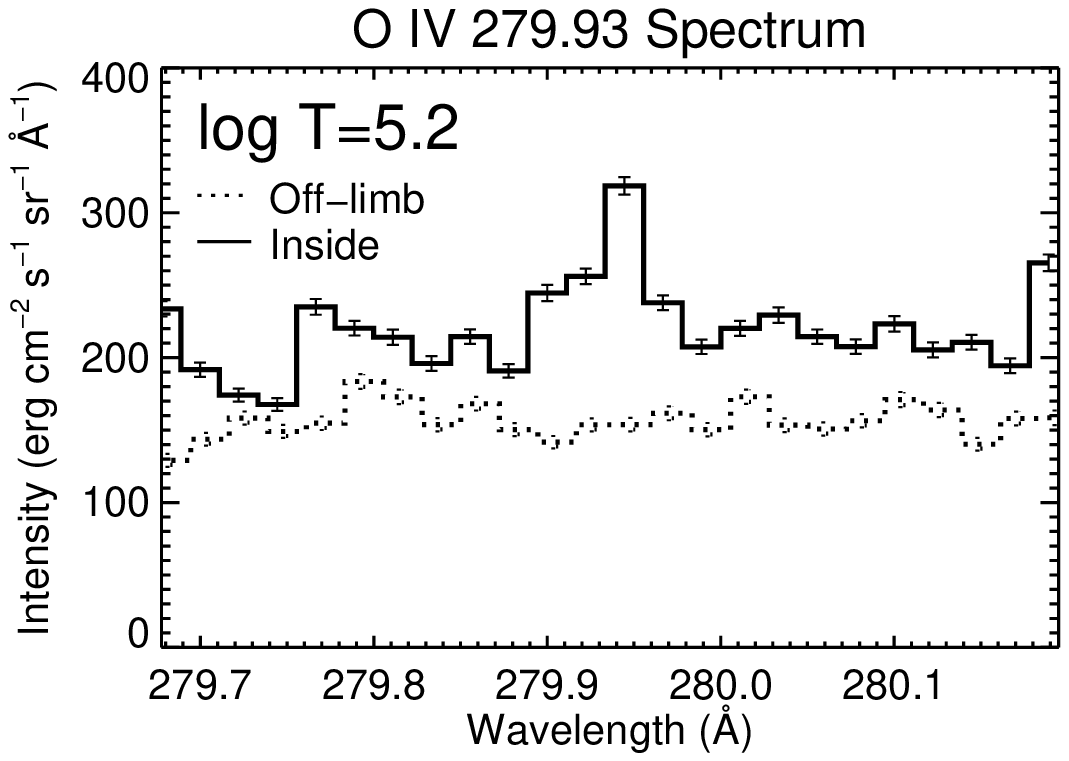}
  \includegraphics[width=7.0cm,clip]{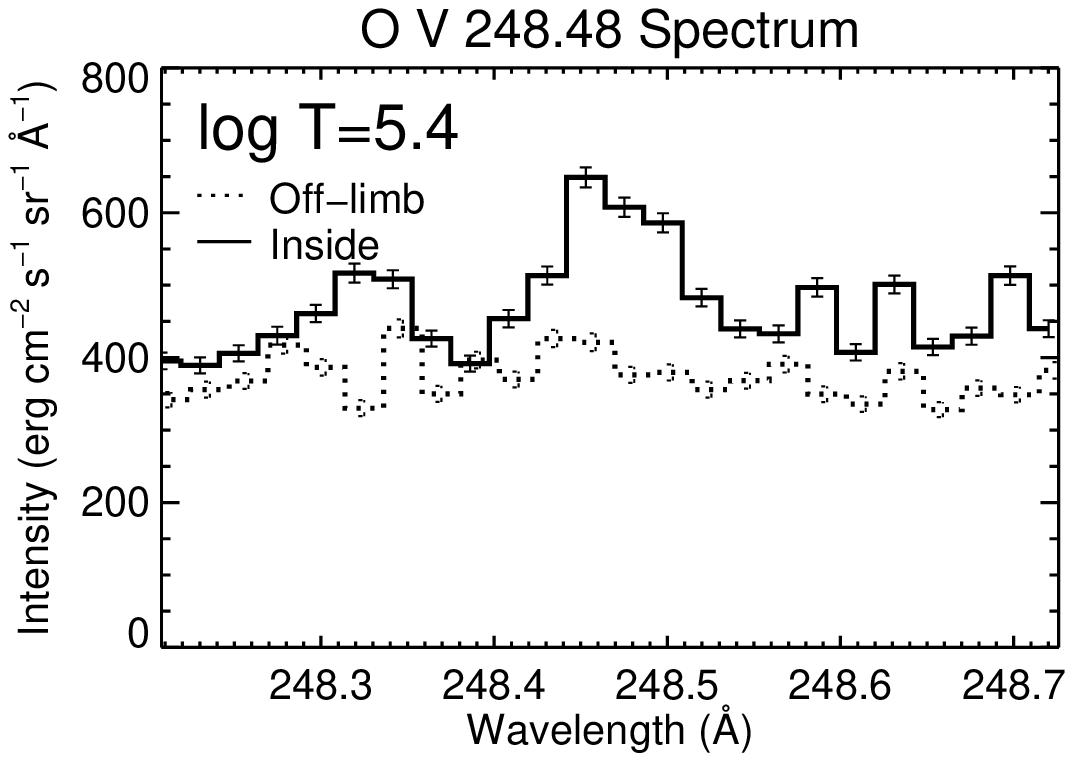}
  \includegraphics[width=7.0cm,clip]{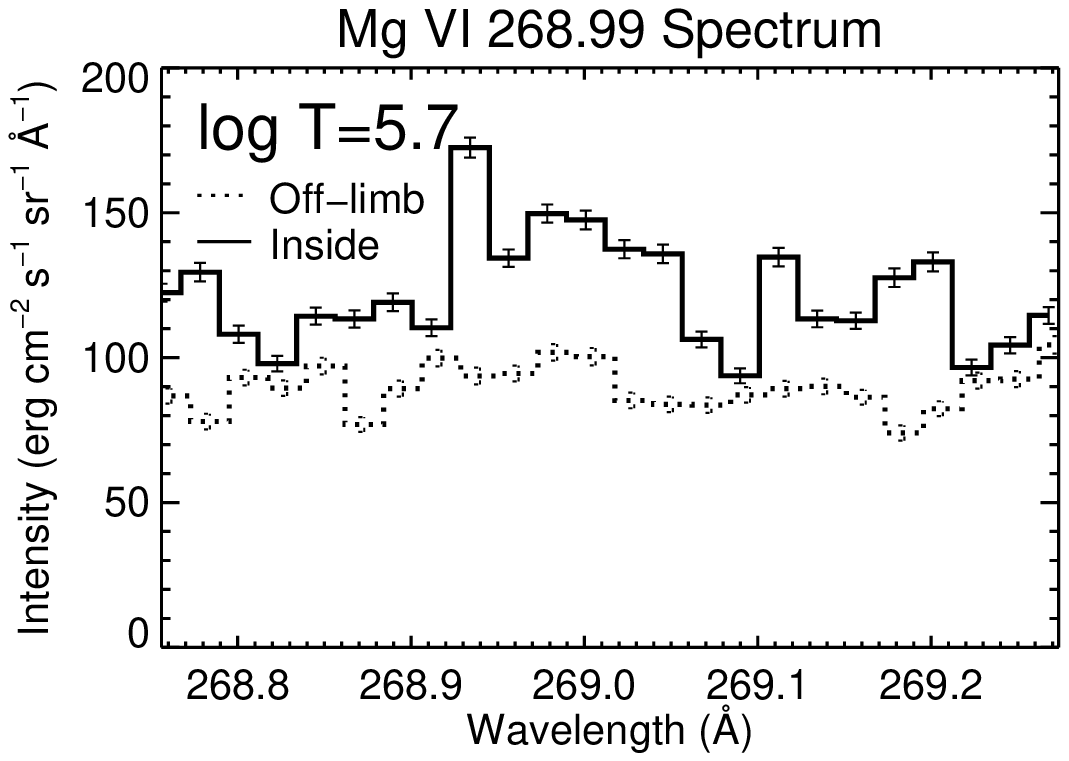}
  \includegraphics[width=7.0cm,clip]{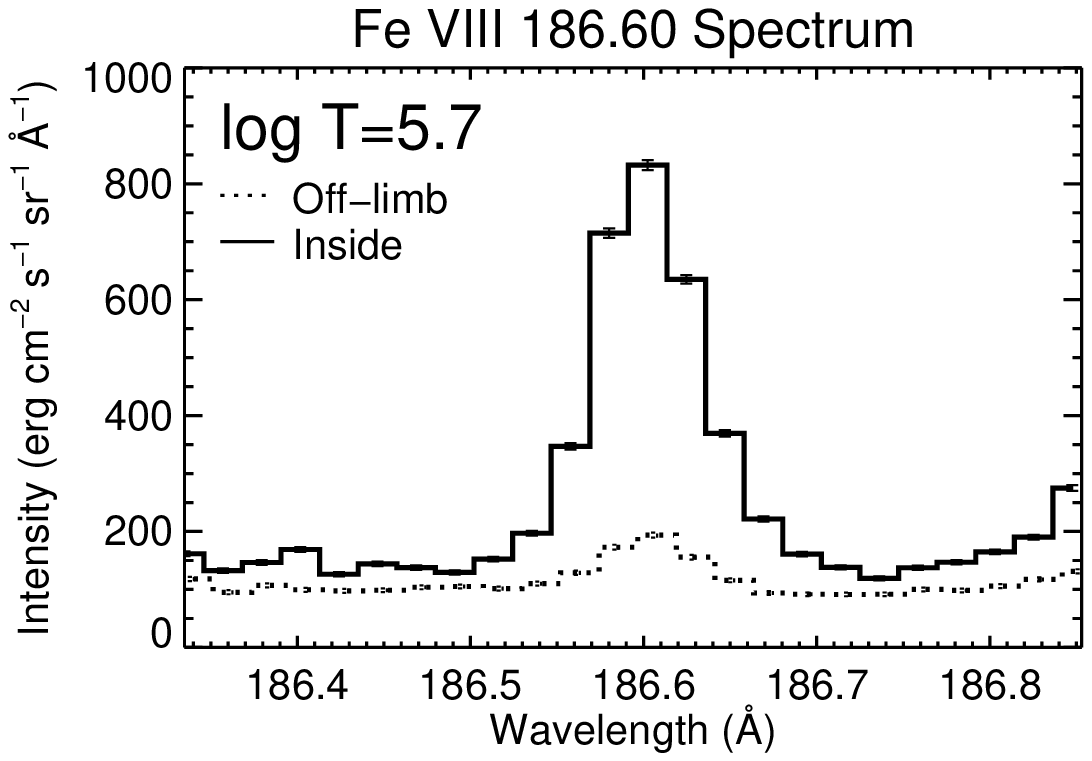}
  \includegraphics[width=7.0cm,clip]{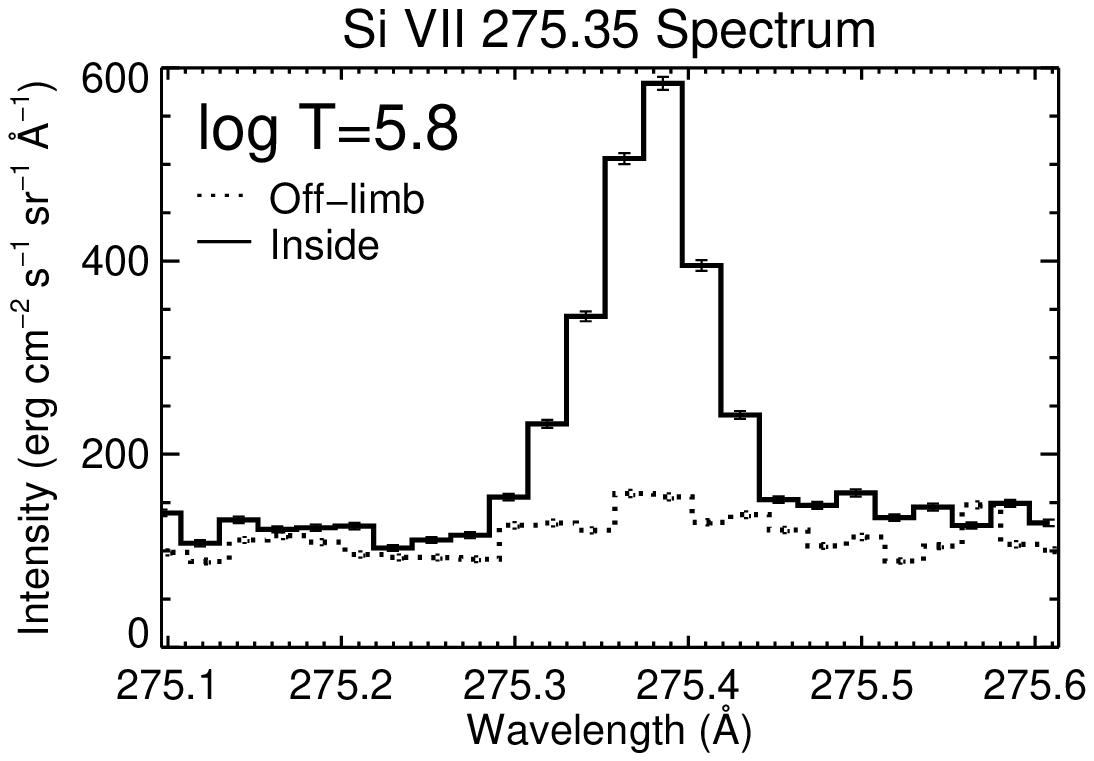}
  \includegraphics[width=7.0cm,clip]{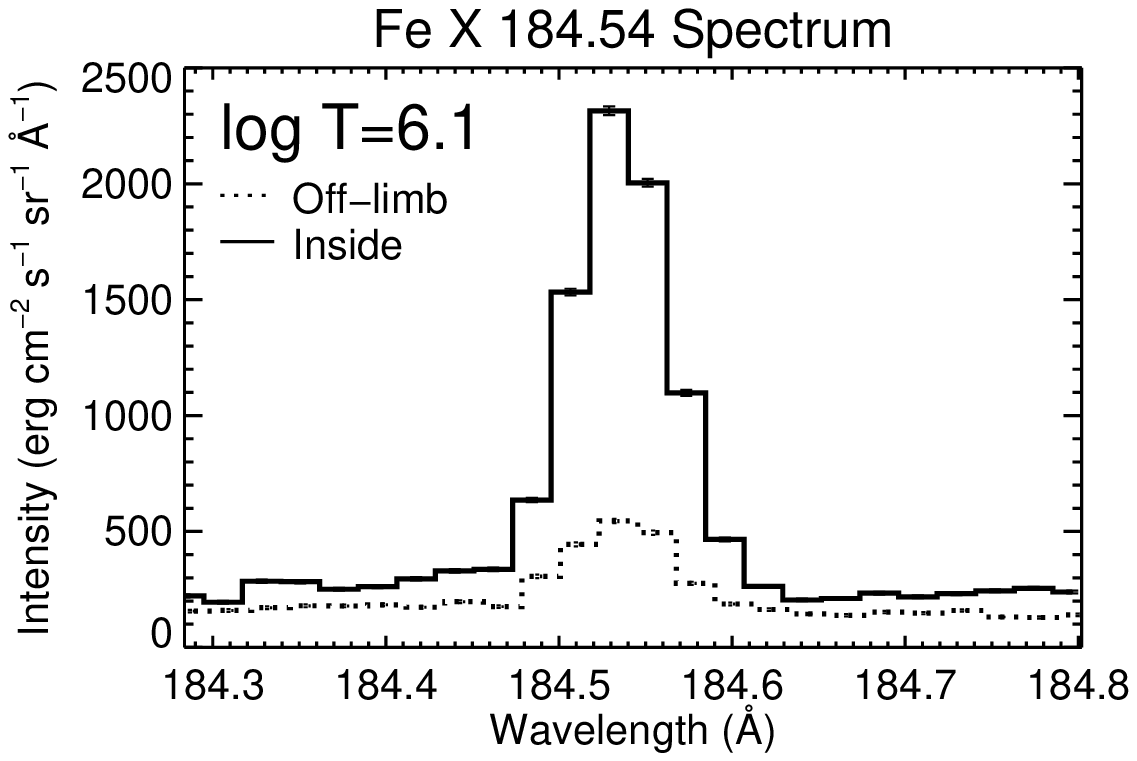}
  \includegraphics[width=7.0cm,clip]{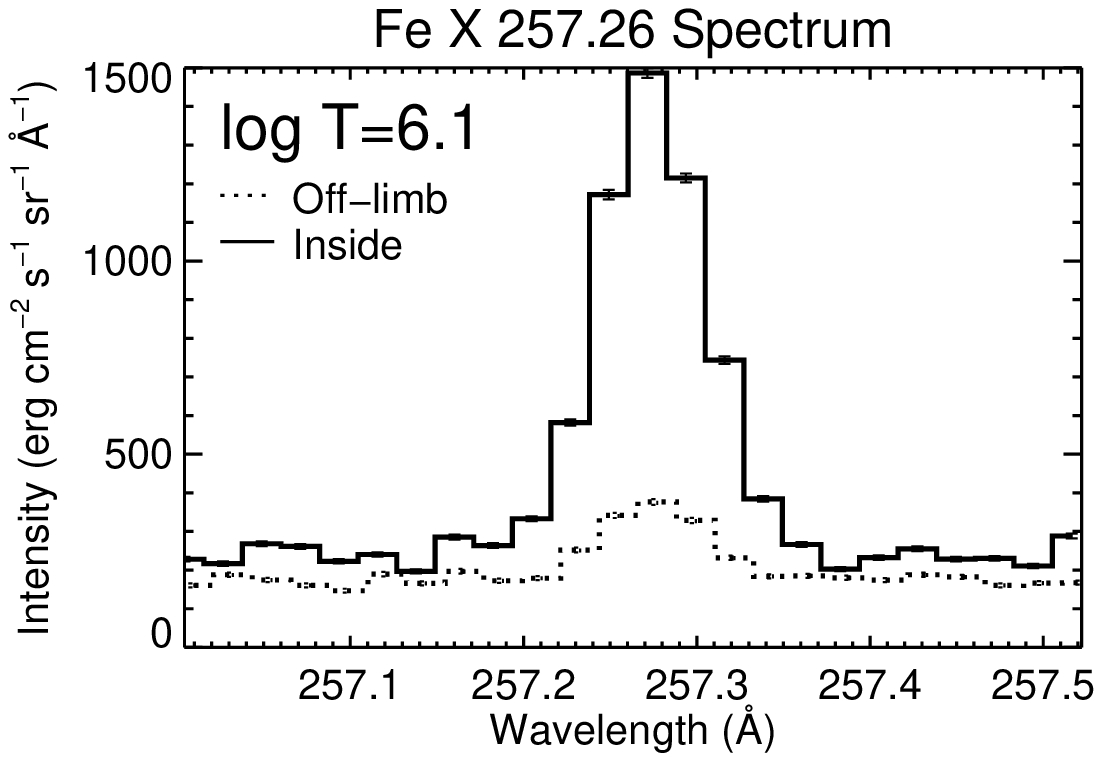}
  \caption{Line profiles on the disk ({solid}) and off the limb ({dotted}).  
The number in the upper-left corner of each panel is the logarithmic formation 
temperature of the emission line.  Error bars include the photon noise and 
the uncertainty in the CCD pedestal and dark current.}
  \label{fig:cal_lp_ex}  
\end{figure}
\addtocounter{figure}{-1}

\begin{figure}
  \centering
  \includegraphics[width=7.0cm,clip]{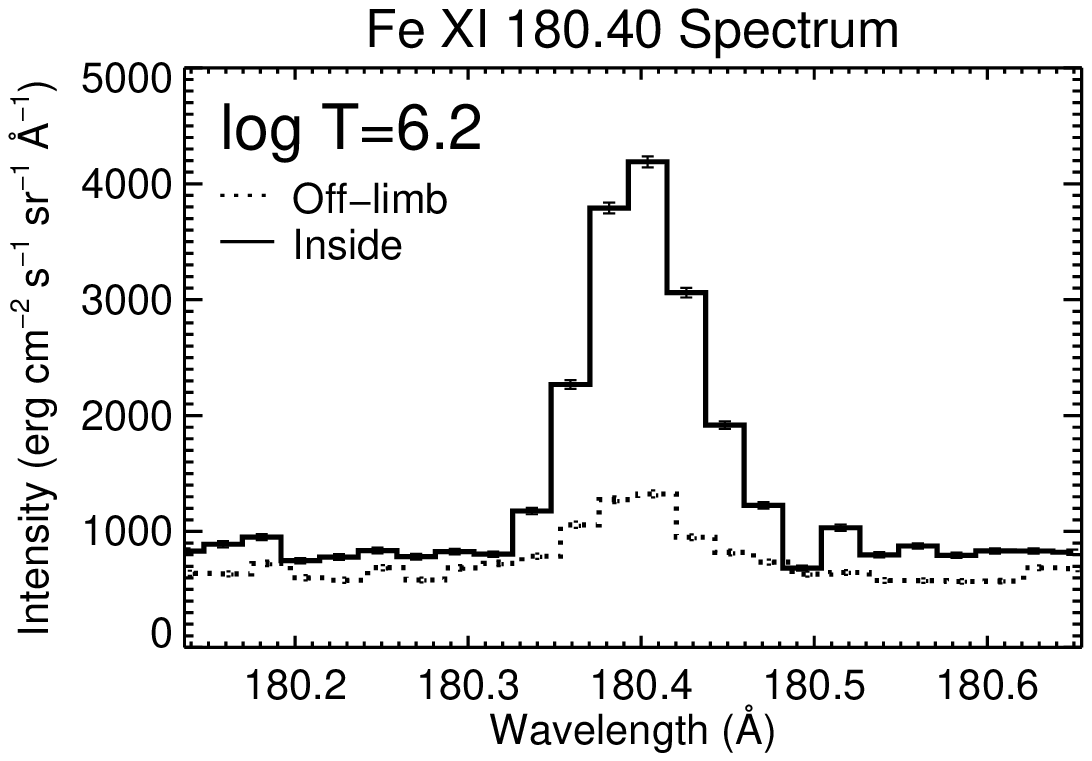}
  \includegraphics[width=7.0cm,clip]{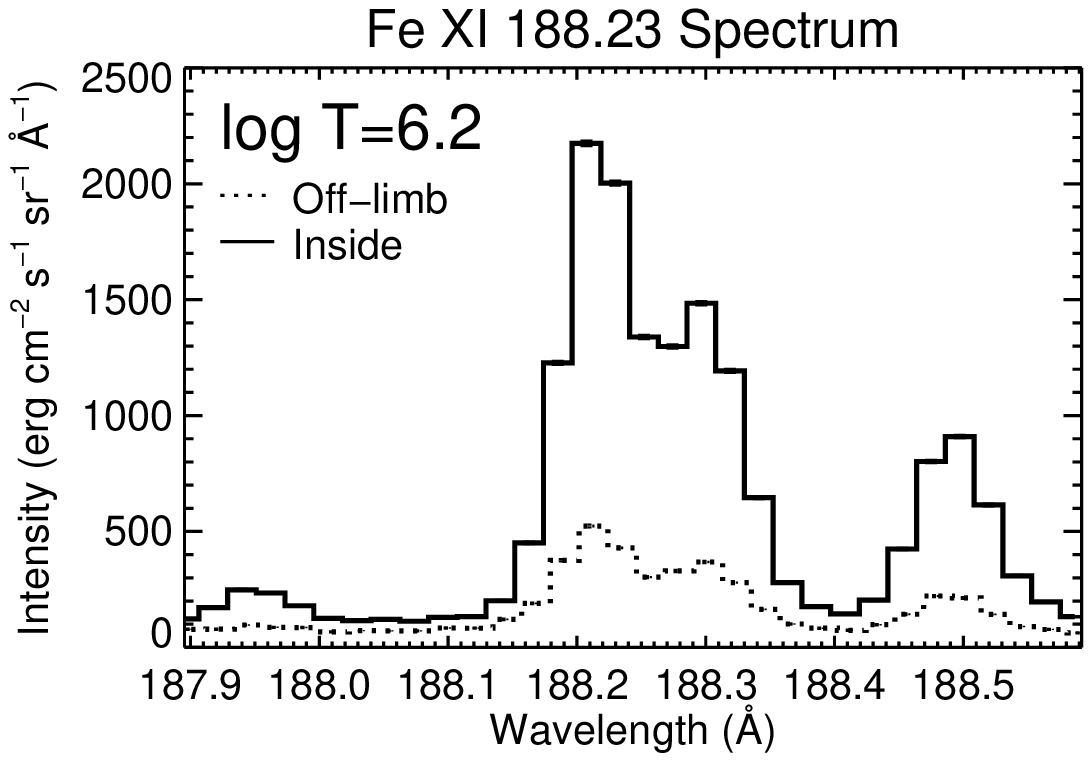}
  \includegraphics[width=7.0cm,clip]{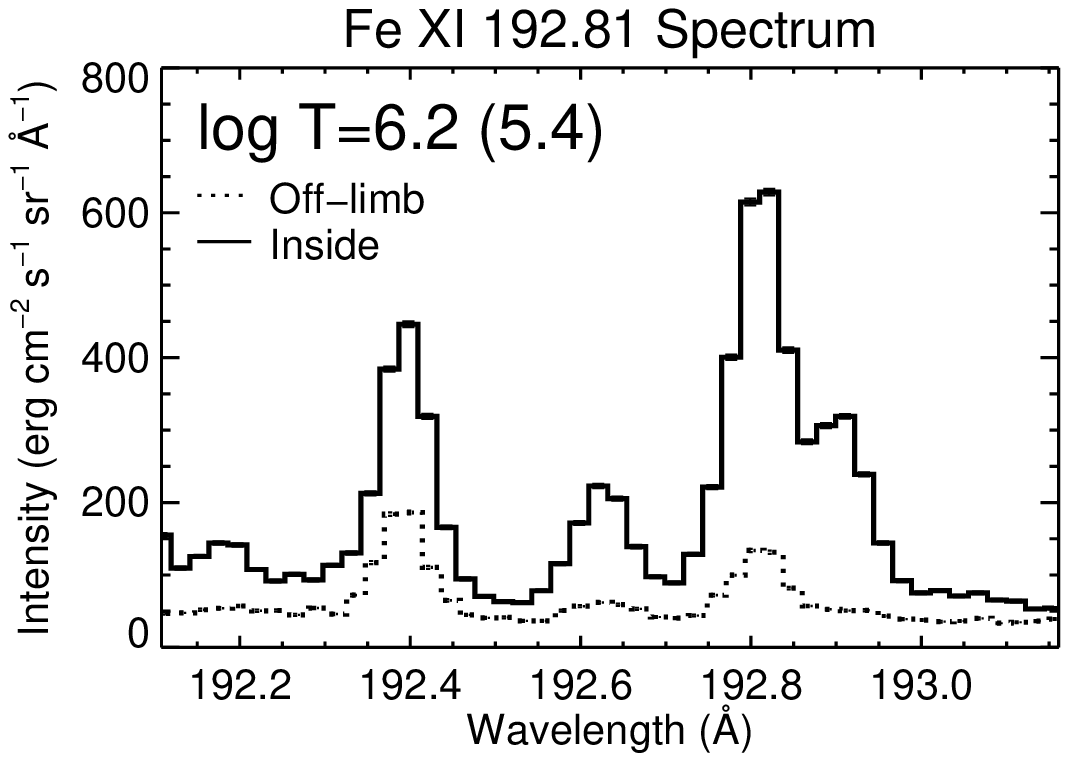}
  \includegraphics[width=7.0cm,clip]{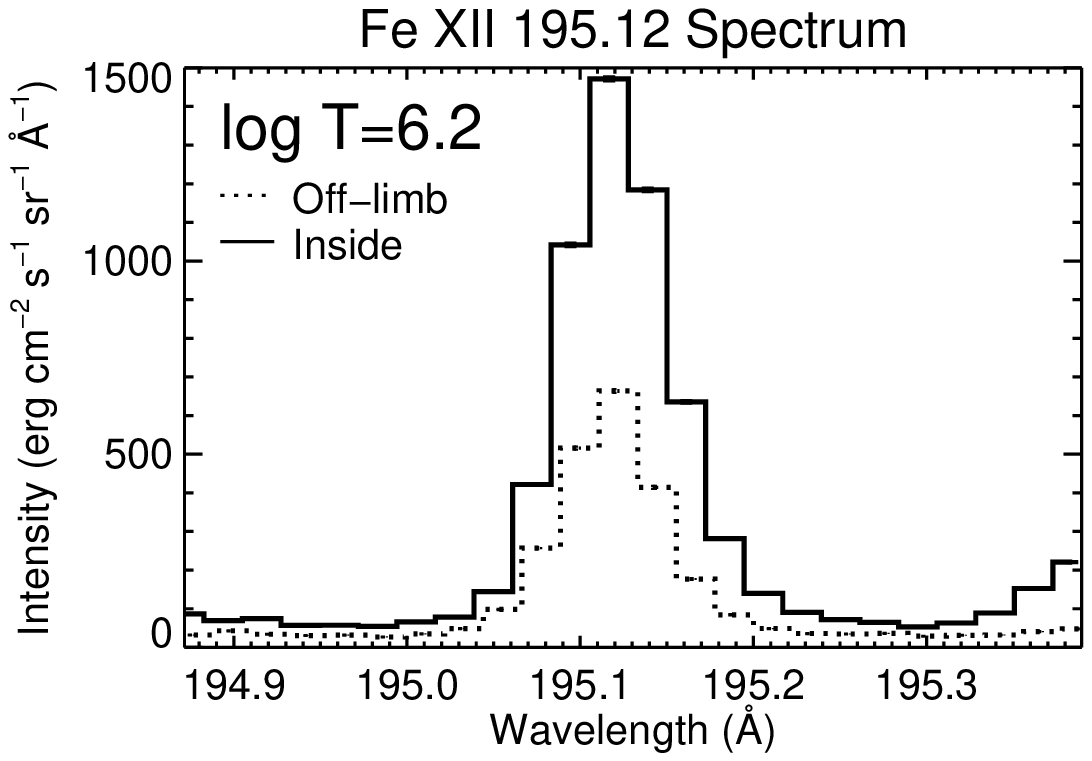}
  \includegraphics[width=7.0cm,clip]{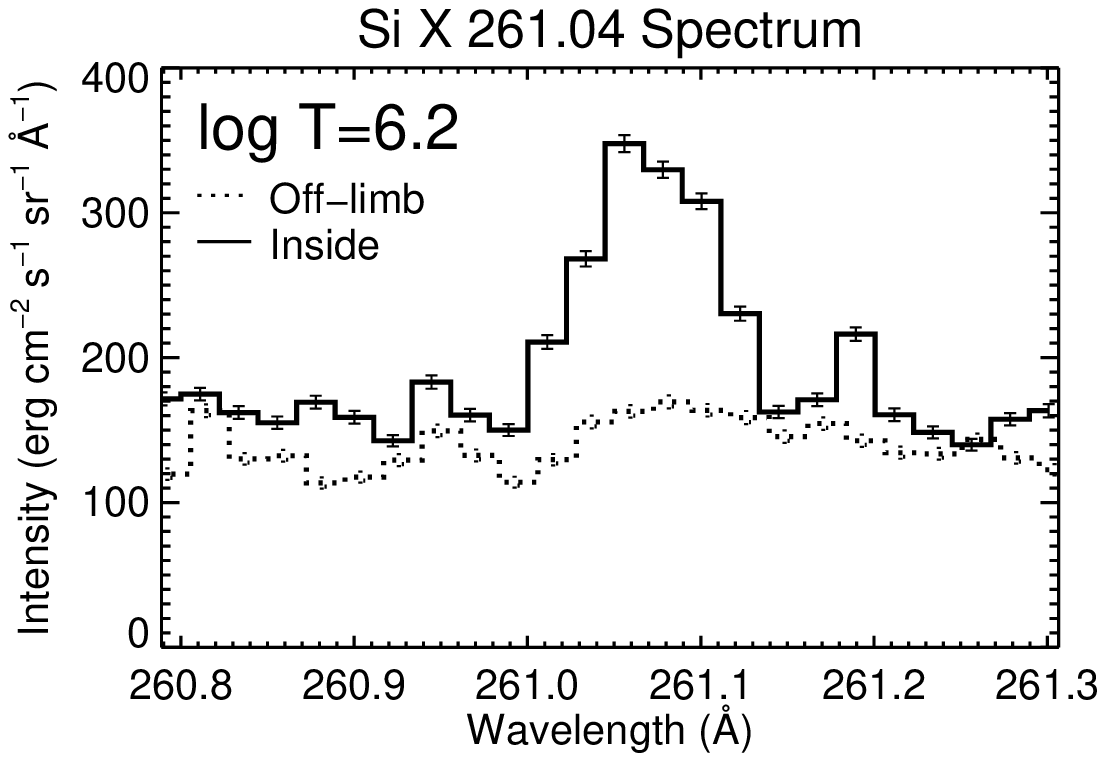}
  \includegraphics[width=7.0cm,clip]{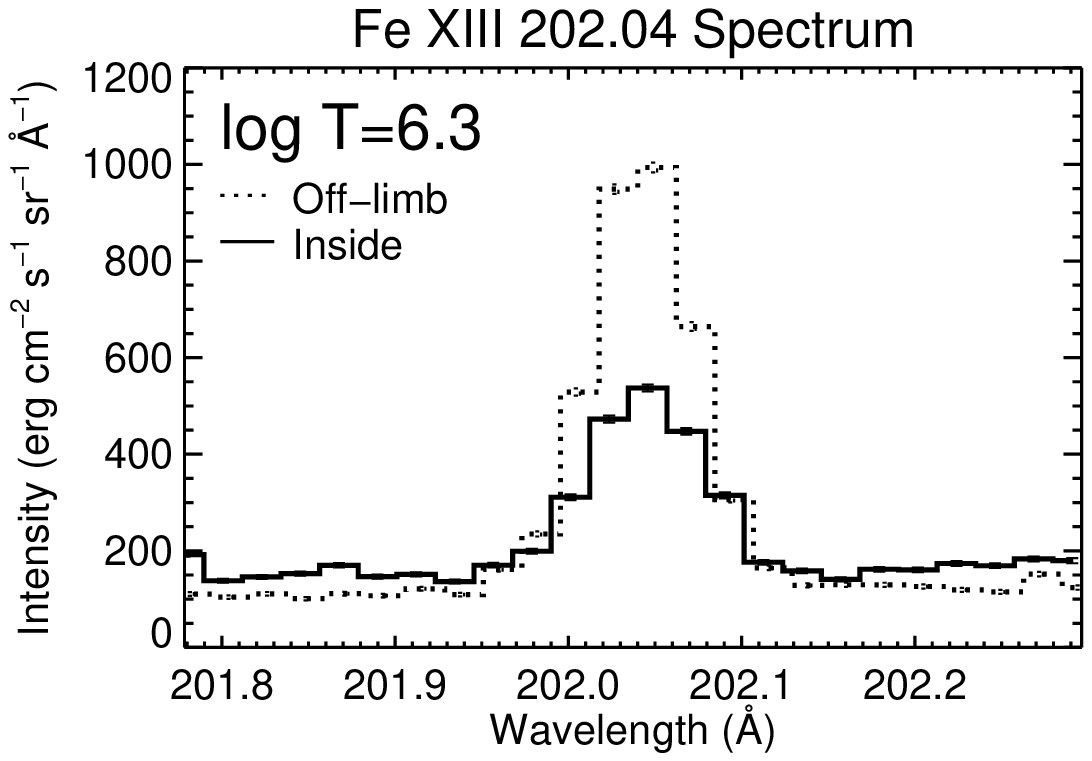}
  \includegraphics[width=7.0cm,clip]{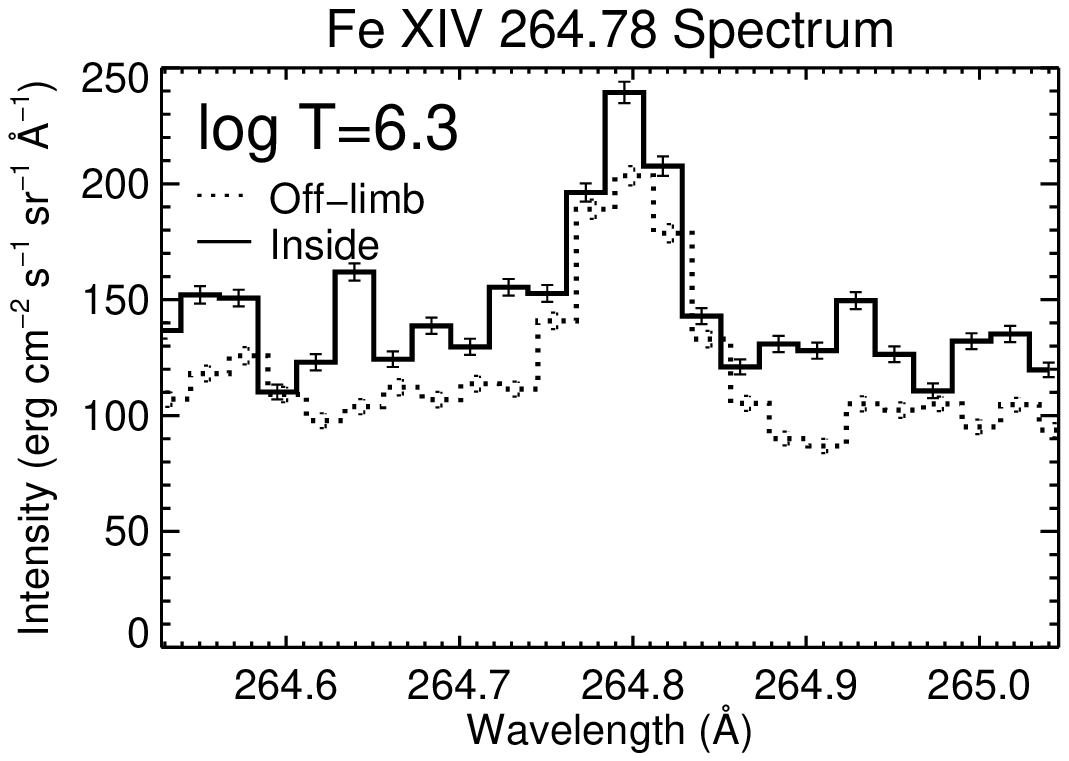}
  \includegraphics[width=7.0cm,clip]{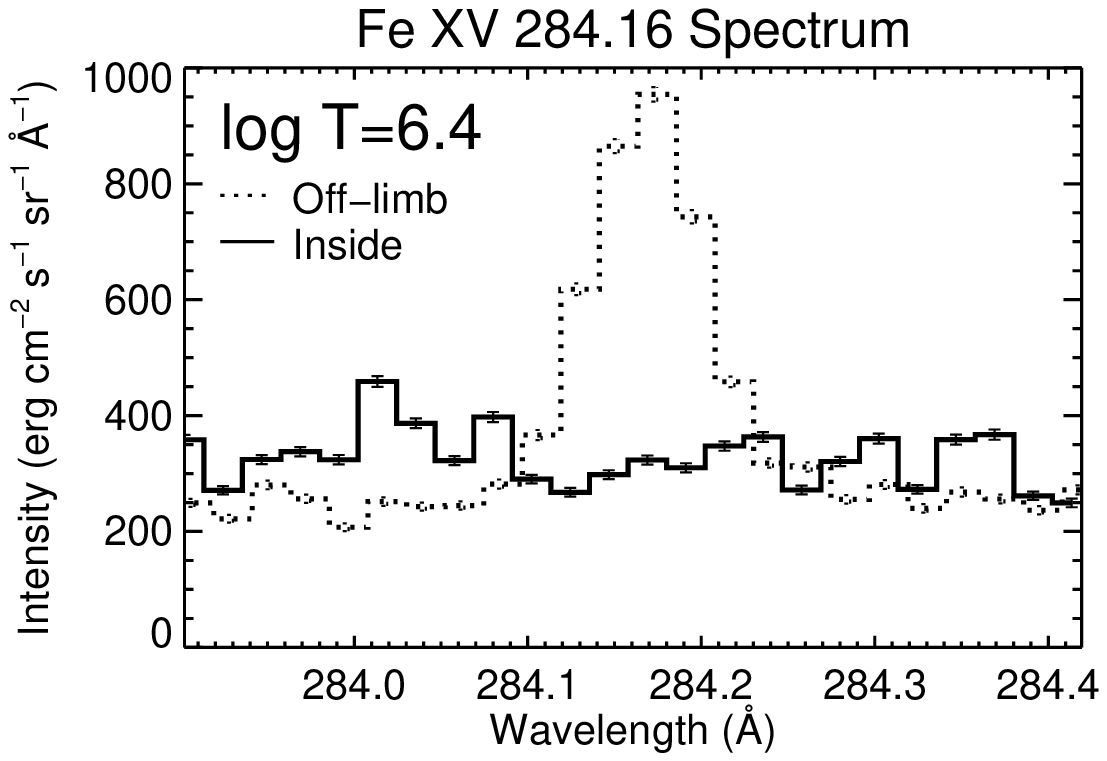}
  \caption{{Continued.}}
\end{figure}

He \textsc{ii} $256.32${\AA} is one of the strongest emission 
lines in EIS spectra and the only one with a formation temperature below 
$\logt = 5.0$.  The emission is very weak above the limb, indicating that 
the line originates from the bottom of the corona or lower.  
The solid line 
profile He \textsc{ii} $256.32${\AA} reveals a long enhanced red wing
contributed from Si \textsc{x} $256.37${\AA}. This blending greatly
complicates the analysis of He \textsc{ii} $256.32${\AA}.  Ideally, 
the Si \textsc{x} can be removed by referring Si \textsc{x} $261.04${\AA}:
Because both lines share the same upper transition level, 
this line pair has a constant intensity ratio 
($I_{256.37} / I_{261.04} = 1.25$; CHIANTI ver.\ 7, \citeauthor{dere1997}\
\citeyear{dere1997}; \citeauthor{landi2012}\ \citeyear{landi2012}).
The intensity ratio could be measured above the limb, 
where He \textsc{ii} becomes much weaker than it was inside the solar disk.  
However, as seen from the 
off-limb spectrum ({dotted histogram} in Fig.~\ref{fig:cal_lp_ex}) 
of Si \textsc{x} $261.04${\AA},
this line was too weak ({i.e.}, noisy) to be used as a 
reference emission line; thus, it was excluded in
the present analysis.

The analyzed EIS data include two oxygen emission lines: O \textsc{iv} 
$279.93${\AA} ($\logt = 5.2$) and O \textsc{v} $248.48${\AA} ($\logt = 5.4$).  
In previous reports, the transition region lines around 
$\logt \simeq 5.0$--$5.5$ were redshifted by up to $\sim 10 \, \kmpers$ at 
the disk center \citep{chae1998doppler,peterjudge1999,teriaca1999}.
Therefore, to check whether our results are consistent with 
previous observations, we could potentially focus on these oxygen lines. 
Since the oxygen emission lines yield much weaker spectra than 
those of other observed emission lines 
({e.g.}, Fe emission lines), we integrated their spectra 
almost along the entire slit ($500''$) at the expense of spatial resolution.
As seen from the spectra in Fig.~\ref{fig:cal_lp_ex}, the integrating over 
$100$ pixels is too coarse for measuring the precise line 
centroid.

The emission lines Fe \textsc{viii} $186.60${\AA} and Si \textsc{vii} 
$275.35${\AA} are strong and well-isolated from the other strong lines; in 
addition, their formation temperatures are similar.  A Ca 
\textsc{xiv} emission line appears near the line centroid of Fe \textsc{viii} 
$186.60${\AA}, but its influence is considered to be very weak in the quiet 
region because of the high formation temperature of Ca \textsc{xiv} 
($\logt=6.7$). Although the formation temperature of
Mg \textsc{vi} $268.99${\AA} is similar to that of 
Fe \textsc{viii} and Si \textsc{vii}, this line was too noisy to achieve a 
precision of several $\kmpers$; thus, it was rejected.

At the longer wavelength side in the spectral window of Fe \textsc{xi} 
$188.21${\AA}/$188.30${\AA}, there is a Fe \textsc{ix} $188.49${\AA} line, 
which is isolated and relatively strong. Using this line, we 
can fill the wide temperature 
gap between Fe \textsc{viii} ($\logt = 5.69$) and Fe 
\textsc{x} ($\logt = 6.04$).

There are two emission lines from Fe \textsc{x} in the analyzed EIS data: 
$184.54${\AA} and $257.26${\AA}.
Neither line is significantly blended by other 
lines near the line center,  although 
a weak line Fe \textsc{xi} $184.41${\AA} exists in
the red wing of Fe \textsc{x} $184.54${\AA}. However,
this line is much weaker than Fe \textsc{x} $184.54${\AA} in the quiet region. 

Fe \textsc{xi} emission lines are included in three spectral windows: 
$180.40${\AA}, $188.21${\AA}, and $192.81${\AA}.  Unfortunately, all
of these lines are significantly blended.  The line center of Fe \textsc{xi} 
$180.40${\AA} is very close to Fe \textsc{x} $180.44${\AA} 
(separation of $\sim 2$ pixels in the EIS CCD).  This emission line is 
density-sensitive and strengthens in regions of high electron density.  
This strengthening may cause a systematic redshift relative to other 
Fe \textsc{xi} emission lines.  The Fe \textsc{xi} $188.21$ line 
is blended with another emission line of comparative strength 
(Fe \textsc{xi} $188.21${\AA}/$188.30${\AA}; see Fig.~\ref{fig:cal_lp_ex}).  
Therefore, we fitted Fe \textsc{xi} $188.21${\AA}/$188.30${\AA} 
by double Gaussians. This fitting is expected to be robust 
because both emission lines are strong and their line profiles usually 
feature two distinct peaks.  The third emission line Fe \textsc{xi} 
$192.81${\AA} is significantly blended by the transition region lines 
O \textsc{v} $192.90${\AA} in the quiet region; hence, it was not used.

The emission lines Fe \textsc{xii} $192.39${\AA} and $195.12${\AA} are 
both strong and suitable for analyzing the quiet region.  Unfortunately, 
Fe \textsc{xii} $195.12${\AA} is blended by Fe \textsc{xii} $195.18${\AA} 
and the line ratio $195.18${\AA}/$195.12${\AA} is sensitive to the electron 
density.  Therefore, this emission line will shift toward longer wavelengths 
({i.e.,} redshift) as the density increases. This is especially in 
active regions and at bright points where the electron density is 
one order of magnitude higher than that in the quiet regions.

The only strong emission line from Fe \textsc{xiii} in our EIS study,
Fe \textsc{xiii} $202.04${\AA} is recognized as a 
clean line with no significant blends.  Different from emission lines with 
lower formation temperature, the off-limb spectrum of Fe \textsc{xiii} 
$202.04${\AA} is approximately twice as strong as the disk spectrum
(see Fig.~\ref{fig:cal_lp_ex}), as expected from the limb brightening effect. 

Emission lines Fe \textsc{xiv} $264.78${\AA} and Fe \textsc{xv} $284.16${\AA} 
were very weak in the quiet region even after an exposure time of 
$120 \, \mathrm{s}$.  The off-limb and disk spectral behaviors of 
Fe \textsc{xv} mimics those of Fe \textsc{xiii},  but those of 
Fe \textsc{xiv} are very different, probably because the latter is influenced 
by the nearby emission line Fe \textsc{xi} $264.77${\AA}.  In the quiet region,
where the average temperature is lower than that in active regions,
Fe \textsc{xi} could make a stronger contribution than Fe \textsc{xiv}. 
Fe \textsc{xv} might be similarly affected in the quiet region.  
Therefore, the Fe \textsc{xiv} and Fe \textsc{xv} emission lines 
were excluded as they might have yielded improper results.

%\bibliography{apj-jour,bib_nk_a,bib_nk_b,bib_nk_c,bib_nk_d,bib_nk_e,bib_nk_g,bib_nk_h,bib_nk_i,bib_nk_k,bib_nk_l,bib_nk_m,bib_nk_o,bib_nk_p,bib_nk_r,bib_nk_s,bib_nk_t,bib_nk_u,bib_nk_w,bib_nk_y}
\bibliography{apj-jour,bib_nk}

\end{document}